\documentclass[preprintnumbers,nofootinbib]{revtex4}
\usepackage{eurosym}
\usepackage{graphicx}
\usepackage{amsmath,amssymb,subfigure}
\usepackage{color}

\def\bc{\begin{center}}
\def\ec{\end{center}}

\def\beq{\begin{equation}}
\def\eeq{\end{equation}}

\def\bc{\begin{center}}
\def\ec{\end{center}}

\def\beq{\begin{equation}}
\def\eeq{\end{equation}}

\begin{document}

\title{Few-body systems in condensed matter physics. }
\author{Roman Ya. Kezerashvili$^{1,2}$}
\affiliation{\mbox{$^{1}$Physics Department, New York
City College
of Technology, The City University of New York,} \\
Brooklyn, NY 11201, USA \\
\mbox{$^{2}$The Graduate School and University Center, The
City University of New York,} \\
New York, NY 10016, USA \\
}

\begin{abstract}
This review focuses on the studies and computations of few-body systems of
electrons and holes in condensed matter physics. We analyze and illustrate
the application of a variety of methods for description of two- three- and
four-body excitonic complexes such as an exciton, trion and biexciton in
three-, two- and one-dimensional configuration spaces in various types of
materials. We discuss and analyze the contributions made over the years to
understanding how the reduction of dimensionality affects the binding energy
of excitons, trions and biexcitons in bulk and low-dimensional
semiconductors and address the challenges that still remain.
\end{abstract}

\maketitle

\section{\protect\bigskip Introduction}

Over the past 50 years, the physics of few-body systems has received
substantial development. Starting from the mid-1950's, the scientific
community has made great progress and success towards the development of
methods of theoretical physics for the solution of few-body problems in
physics. In 1957, \ Skornyakov and Ter-Martirosyan \cite{TerMartiros} solved
the quantum three-body problem and derived equations for the determination
of the wave function of a system of three identical particles (fermions) in
the limiting case of zero-range forces. The integral equation approach \cite%
{TerMartiros} was generalized by Faddeev \cite{Faddeev} to include finite
and long range interactions. It was shown that the eigenfunctions of the
Hamiltonian of a three-particle system with pair interaction can be
represented in a natural fashion as the sum of three terms, for each of
which there exists a coupled set of equations. It should be noted that the
natural division of the wave function in the three-particle problem into
three terms had also been considered earlier by Eyges \cite{Eyges} and
Gribov \cite{Gribov}. However, by using this division of the wave function
in the three particle problem into three terms Faddeev obtained a set of
integral equations with an unambiguous solution. In the limit of zero range,
one obtains the well-known Skornyakov-Ter-Martirosyan equations \cite%
{TerMartiros}. The introduction of separable potentials allowed us to turn
the problem of finding the amplitudes to that of solving a one-dimensional
integral equation. The complete mathematical study and investigation of
three-body problem in discrete and continuum spectra was done by L. D.
Faddeev in Ref. \cite{FaddeevTrudi}. "Those days Faddeev was already a
prominent figure in quantum physics" \cite{Efimov2019}. The studies of
three-body physics led to discovery of the Efimov's effect \cite{Efimov1970,
Efimov1970PL}. When Efimov in 1970 discussed this phenomenon with Faddeev,
"Faddeev was very surprised. In a few days he [Faddeev] called me [Efimov]
and said he confirmed my results using his own method" \cite{Efimov2019}.%
\footnote{%
In the summer of 2016 at the EFB23 conference in Aarhus, Denmark, following
the inaugural ceremony establishing the Faddeev Medal, I spoke with L. D.
Faddeev and mentioned that I would like to nominate Vitaly Efimov for this
distinguished award. A smile touched Faddeev's face and he said simply,
"great choice". In 2018, Vitaly Efimov and Rudolf Grimm became joint
recipients of the Faddeev Medal for the theoretical prediction and
ground-breaking experimental confirmation of the Efimov effect.}

As for the four-body system, Faddeev's idea of an explicit cluster-channel
separation was completely elaborated by Yakubovsky in 1967 \cite{Yakubovsky}%
. Merkuriev, Gignoux and Laverne \cite{Merkuriev} in 1976 found and studied
the asymptotic boundary conditions that are needed in configuration space in
order to find unique physical solutions corresponding to various scattering
processes which results in the Faddeev differential equations \cite{MF85}.
Therefore, the importance of the Faddeev method development in the
coordinate representation was demonstrated. %Over the past 50 years, the
%physics of few-body systems has received substantial development. It
%currently includes the problems of traditional nuclear and hypernuclear
%physics, quark physics, atomic physics and quantum chemistry physics (the
%structures of molecules of three or more particles).
Formulation of Faddeev integral equations has triggered the development of
other approaches for solutions of few-body problems in physics. The general
approach for solutions of these problems is based on the use of modelless
methods for studying the dynamics of few-body systems in discrete and
continuum spectra. Currently among the most powerful approaches are the
method of hyperspherical harmonics (HH), the variational method in the
harmonic-oscillator basis and the variational method complemented with the
use of explicitly correlated Gaussian basis functions \cite{Varga1, BubinRMP}%
. The hyperspherical harmonics method occupies an important place.
Hyperradial equations are obtained from the three-particle Schr\"{o}dinger
equation by considering the orthonormality of HH. The analogous equations
were obtained in the early works of Morpurgo \cite{Morpurgo}, Delves \cite%
{Delves, Delves2, Delves3}, and Smith \cite{Smith}, but the method became
particularly popular after the works of Simonov \cite{Simonov}, and Badalyan
and Simonov \cite{Simonov2}. Despite its conceptual simplicity, the method
of hyperspherical harmonics offers great flexibility, high accuracy, and can
be used to study diverse quantum systems, ranging from small atoms and
molecules to light nuclei, hadrons, quantum dots, and Efimov systems. The
basic theoretical foundations and details of this method are discussed in
monographs \cite{Avery, Jibuti2}. Over the past 50 years, the physics of
few-body systems has received substantial development. It currently includes
the problems of traditional nuclear and hypernuclear physics, quark physics,
atomic physics and quantum chemistry (the structures of molecules of three
or more particles). The rapid development of the theory stimulated
experimental studies of various properties of few-body systems in different
areas of physics.

This review presents a variety of approaches for the description of few-body
systems of electrons and holes, namely the two-body (exciton), three-body
(trions, or charged excitons), and four-body (biexciton) systems collectively
known as excitonic complexes, in condensed matter physics. These excitonic
complexes were experimentally observed in three-dimensional (3D) bulk
materials, two-dimensional (2D) novel layered materials and one-dimensional
(1D) materials. Although the excitonic complexes like excitons, trions,
biexcitons in condensed matter physics are very similar to the two- three-
and four-body bound systems in atomic and nuclear physics, there are major
differences: i. Excitonic complexes are excited in bulk materials as 3D
systems, in novel atomically thin materials they are 2D systems and in
nanowires, nanorods and nanotubes these complexes are considered as 1D
systems; ii. The reduction of dimensionality itself necessitates a change to
the formalism, and requires the modification of the bare Coulomb potential
to account for non-local screening effects. The screening effects, resulting
from the host lattice, make the Coulomb force between charge carriers much
weaker than in atomic systems; iii. Band effects make the effective masses
of the electrons and holes smaller than the bare electron mass.

In this review we discuss and focus on the application of a variety of
theoretical approaches for the description of two- three- and four-body
excitonic complexes in 3D, 2D and 1D configuration spaces in condensed
matter physics, as well as how the reduction of dimensionality affects the
binding energy of excitons, trions and biexcitons in bulk and
low-dimensional semiconductors. The excitons, trions and biexcitons in 3D,
2D and 1D configuration spaces are discussed in Sec. II, III and IV,
respectively. Conclusions follow in Sec. V.

\section{Two Body Problem$-$Excitons}

An exciton is an elementary excitation in condensed matter created when a
conduction band electron and a valence band hole form a bound state due to
the Coulomb attraction. It can be formed by absorption of a photon in a
semiconductor by exciting the electron from the valence band into the
conduction band. The exciton is an electrically neutral quasiparticle which
can transport energy without transporting a net electric charge. The
electron and hole may have either parallel or anti-parallel spins giving
rise to exciton fine structure when the spins are coupled by the exchange
interaction.

\subsection{3D excitons}

There are two types of excitons: the Mott--Wannier \cite{WannierM} and
Frenkel \cite{Frenkel} excitons. The Mott--Wannier exciton presents a
two-body system and can be treated as an exotic atomic state akin to that of
a hydrogen atom. However, the effective masses of the excited electron and
hole are comparable, and the screening of the Coulomb attraction leads to a
much smaller binding energy and larger radius than the hydrogen atom. The
recombination of the electron and hole, i.e. the decay of the exciton, is
limited by resonance stabilization. Frenkel excitons were introduced by
Frenkel \cite{Frenkel} and are formed in materials with relatively small
dielectric constants, which results in a relatively strong Coulomb
attraction, leading to excitons of relatively small size, of the same order
as the size of the unit cell. The Mott--Wannier excitons \cite{WannierM} are
formed in semiconductors with relatively large dielectric constants and
small band gaps. As a result of the weaker electron-hole attraction due to
the stronger screening, the radius of the Mott-Wannier exciton exceeds the
lattice spacing. The screening of the Coulomb attraction in bulk materials
(3D materials) is a result of the macroscopic polarization induced by a
point charge surrounded by a 3D dielectric medium (Fig. 1). The electric
field at a point $\mathbf{r}$ from the charge is the sum of the external
field produced by the electron, $ke\mathbf{r}/r^{3}$, and the induced field
due to the polarization of the medium. This charge distribution produces a
field of the same functional form $ke\mathbf{r}/\varepsilon r^{3}$ and the
screening is given by a simple multiplicative renormalization through the
dielectric constant $\varepsilon $. Therefore, the binding energy of the
Mott-Wannier excitons are obtained by the solution of the Schr\"{o}dinger
equation with the Coulomb potential renormalized by the dielectric constant $%
\varepsilon $ only$.$ Consequently the eigenstates energies of excitons have
a Rydberg series structure. \ However, the effective Bohr radius for an
exciton, may be much larger than the Bohr radius of the hydrogen atom, and
the exciton binding energy is much smaller than the binding energy of the
hydrogen atom. The two body Schr\"{o}dinger equation with the screening
Coulomb interaction $ke/\varepsilon r^{2}$ perfectly describes all
excitonic effects in 3D materials.

\begin{figure}[t]
\centering
\par
\includegraphics[width=0.30\textwidth]{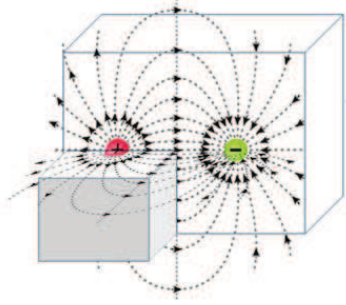}
\caption{ Electric field lines for two interacting particles in a
uniform dielectric environment in 3D materials. }
\label{F1}
\end{figure}

\subsection{2D excitons}

Since the experimental discovery of highly conductive graphene monolayers in
2004~\cite{Novoselov}, the field of condensed matter physics has seen
explosive growth in theoretical and experimental research in the realm of
two-dimensional materials. The isolation of graphene from bulk crystals of
graphite allowed the identification of just the first member of a family of
2D layered materials, which has grown rapidly over the past ten years and
now includes insulators, semiconductors, semimetals, metals, and
superconductors \cite{Bhimanapati, Novoselov2}. 2D materials are commonly
defined as crystalline materials consisting of a single layer of atoms. Most
often such materials are classified as either 2D allotropes of various
elements or compounds consisting of two or more ionically/covalently bonded
elements. 2D materials have, within just one decade, reshaped many
disciplines of modern science, both through intensive experimental and
theoretical studies of their properties, as well as by providing a rich
platform for further exploration of previously-known and newly-emerging
exotic physical phenomena. These materials are crystalline solids with a
high ratio between their lateral size and thickness \cite{Velicky}. In these
layered materials, known also as van der Waals materials, the atomic
organization and bond strength in the 2D plane are typically much stronger
than in the third dimension (out-of-plane), where they are bonded together
by weak van der Waals' interaction \cite{Novoselov2, Jariwala}. Today,
research in the field is primarily focused on graphene, transition metal
dichalcogenides (TMDCs) and other emerging 2D materials beyond graphene such
as phosphorene and transition metal trichalcogenides (TMTCs), which are the
anisotropic semiconductors, and Xenes. These nanomaterials are essential for
the next generation of devices in tunable optoelectronics, sensing, and
photovoltaics. The gapless nature of graphene makes it less ideal for the
study of optical phenomena in 2D crystals \cite{Novoselov3}. In terms of
materials, the most well studied 2D materials beyond graphene are the
semiconducting transition metal dichalcogenides with the chemical formula MX$%
_{2}$, where M denotes a transition metal M = Mo, W, and X denotes a
chalcogenide X = S, Se, or Te \cite{Kormanyos}, transition metal
trichalcogenides \cite{Joshua} and phosphorene \cite{Li1, Li2, Warren2015}.
In recent years, experimental success was achieved in isolating stable
monolayers of 2D insulators such as hexagonal boron nitride (h-BN)~\cite%
{Dean2010}. It is often the case that any study of 2D materials includes the
use of h-BN as a substrate or spacer.

\bigskip Another recent addition to the 2D universe are the buckled 2D
materials \cite{Matthes2014} collectively referred to as Xenes \cite%
{Molle2017, Brunetti2018}: silicene (Si) \cite{Falko2012}, germanene (Ge)
\cite{Davila}, stanene (Sn) \cite{Zhu2014}, and borophene (B) \cite%
{Mannix2015}. In the buckled 2D materials, the triangular sublattices of the
honeycomb structure are vertically offset from each other by an amount which
is small compared to the 2D bond length. These direct band gap materials
exhibit a Dirac cone near the K/K$^{\prime }$ points, but have a non-zero
gap with a parabolic dispersion in the immediate vicinity of the K/K$%
^{\prime }$ points, making them an intriguing counterpart to gapless
graphene. Most interestingly, the vertical offset between the sublattices
means that the band-gap can be tuned by applying an external electric field,
leading to, among other things, a large and externally tunable exciton
binding energy. In Fig. 2, we schematically illustrate an incomplete zoo of
some of the important members of this 2D family.

\begin{figure}[t]
\centering
\par
\includegraphics[width=0.58\textwidth]{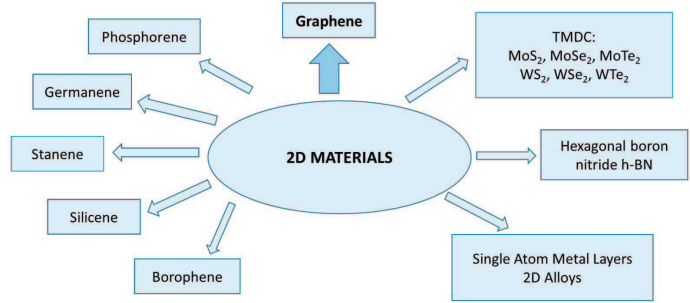}
\caption{Schematic of a zoo of some of the important members of the 2D
family.}
\label{F2}
\end{figure}

New 2D materials offer reasonable flexibility in terms of tailoring their
electronic and optical properties. The emergence of each new material brings
excitement and puzzlement towards their characterization and physical
properties, while expanding the tool set with which scientists may combine
these materials in novel and useful ways. Together, the zoo of 2D materials
offers a truly unique and exciting platform for creating novel
heterostructures with unique properties by stacking the aforementioned 2D
materials in different ways, which the authors of Ref. \cite{Geim2014}
notably described as analogous to building with lego bricks. The properties
of these materials are usually distinctly different from those of their 3D
counterparts, and furthermore, the properties of these materials can
drastically change even when transition from a single, isolated monolayer
to two monolayers separated by a dielectric. An important characteristic of
the 2D materials is the weak and highly non-local way in which they screen
electric fields \cite{Rytova, Keldysh}.

\begin{figure}[t]
\centering
\par
\includegraphics[width=0.30\textwidth]{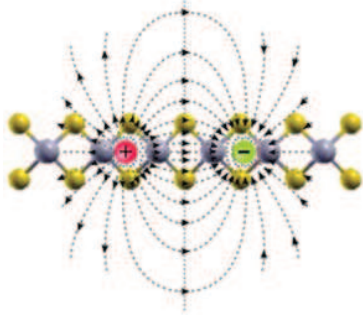}
\caption{ Electric field lines for two interacting particles in 2D
materials.}
\label{F3}
\end{figure}

The reduction of dimensionality has a strong influence on a two-body system
and its effect is twofold: it decreases the kinetic energy and potential
energy of interacting charge carriers. In the two-body problems if the
interaction is described by the Coulomb potential and the dielectric
environment is homogenous, but the electron and hole are constrained to move
in a plane, the reduction of dimensionality affects only the kinetic energy
of the system and one can observe that the spectrum of energy is changed
from $E_{3D}\sim 2\varepsilon \hslash ^{2}/n^{2}$ ($n=1,2,...,$ Rydberg
series) in the 3D case, to $E_{2D}\sim 2\varepsilon \hslash ^{2}/(n-1/2)^{2}$
in the 2D case. Therefore, for example, the ground state energy increases by
a factor of 4. Thus, the reduction of dimensionality suppresses the kinetic
energy of the 2D exciton due to the decrease of the degrees of freedom.
However, the reduction of the dimensionality affects on the potential energy
of the electron-hole interaction, while this interaction is still
electromagnetic by nature, it must be modified from the well-known Coulomb
potential. As mentioned above, in 3D the screening is given by a simple
multiplicative renormalization by the dielectric constant $\varepsilon $ due
to the homogeneous dielectric environment. In contrast to the 3D case, in 2D
case the system is polarizable only in the 2D plane and the induced
polarization field is equivalent to the electric field produced by a uniform
charge distribution on a circle of radius $r$, in contrast to the uniform
charge distribution on the sphere of radius $r$ in case of a 3D material.
Therefore, in comparison to the 3D case shown in Fig. 1, the electric field
outside of the 2D plane shown in Fig. 3 does not affect on two particle
interaction, unlike the portion of the electric field lying within the 2D
monolayer. As a consequence this interaction still will be a function of $r$%
, but with a functional form substantially different from the Coulomb
potential. The 2D electromagnetic interaction between two point charges in a
layered dielectric environment was first derived in Ref. \cite{Rytova}, was
independently re-derived over a decade later in Ref. \cite{Keldysh}, and is
now known as the Rytova-Keldysh (RK) potential. This interaction has the
following form:

\begin{equation}
V(r_{eh})=-\frac{\pi ke^{2}}{\left( \varepsilon _{1}+\varepsilon _{2}\right)
\rho _{0}}\left[ H_{0}\left( \frac{r}{\rho _{0}}\right) -Y_{0}\left( \frac{r%
}{\rho _{0}}\right) \right] .  \label{RKeldysh}
\end{equation}%
In Eq. (\ref{RKeldysh}) $r$ is the distance between the electron and hole, $%
k\equiv 1/(4\pi \epsilon _{0})=9\times 10^{9}\ $N$\cdot $m$^{2}$/C$^{2}$, $%
H_{0}(x)$ and $Y_{0}(x)$ are Struve and Bessel functions of the second kind
of order $\nu =0$, respectively, $\varepsilon _{1}$ and $\varepsilon _{2}$
denote the background dielectric constants on either side of the monolayer,
and the screening length $\rho _{0},$ which sets the boundary between the
two distinct asymptotic behaviors of the RK potential for $r\dashrightarrow 0
$ and $r\dashrightarrow \infty $, is defined by $\rho _{0}=2\pi \zeta /\left[
\left( \varepsilon _{1}+\varepsilon _{2}\right) /2\right] $, where $\zeta \ $%
is the 2D polarizability of the material. For $r>>\rho _{0}$ the potential
has the 3D bare Coulomb tail and becomes $-\frac{ke^{2}}{\varepsilon r}$,
while for $r<<\rho _{0}$ it becomes a logarithmic potential: $-\frac{ke^{2}}{%
\varepsilon \rho _{0\text{ }}}\left[ \ln \left( \frac{r}{2\rho _{0}}\right)
+\gamma \right] ,$ where $\gamma $ is the Euler constant. Thus at small
distance the effect of the induced polarization becomes dominant - the $1/r$
singularity is replaced by a weaker logarithmic dependence. The recent
review of dynamical screening in monolayer transition-metal dichalcogenides
is given in Ref. \cite{Dery2019}.

The exciton can be formed in a double layer system when an electron is
confined in one layer, while the hole is located in a parallel layer
separated by a dielectric of a thickness $D.$ Such excitons with spatially
separated electrons and holes are known as indirect or dipolar excitons, and
were first introduced in Ref.~\cite{Nishanov}. In this system the excitons
can have a much longer lifetime than the direct excitons \cite%
{MoskalenkoSnoke}, because the dielectric barrier between the layers reduces
the probability of electron-hole recombination by tunneling. The prediction
of superfluidity and Bose-Einstein condensation ~\cite{Lozovik} of indirect
excitons in semiconductor coupled layers attracted a great interest to this
system~\cite{Snoke, Butov, Combescot}. The theoretical description of an
indirect exciton is a two-body problem in restricted 3D space when electron
and hole can move each in one of the layers, while motion in the third
direction is restricted by the layers separation $D$. To solve this problem
one projects the electron position vector onto the plane with the hole and
the relative position vector between the electron and the hole is $\mathbf{r}%
_{e}-\mathbf{r}_{h}=\mathbf{r}+D\widehat{\mathbf{z}}$, where $\widehat{%
\mathbf{z}}$ is unit vectors, $D$ is the fixed interlayer separation, and $%
\mathbf{r}$ is the separation between the hole and the projection of the
electron position onto the TMDC layer with holes. One can replace the
relative coordinate by $\sqrt{r^{2}+D^{2}}$ and, as a consequence, the
electron-hole potential should be replaced by %$V(\sqrt{r^{2}+D^{2}}).$

\begin{equation}
V(r)=-\frac{\pi ke^{2}}{\left( \varepsilon _{1}+\varepsilon _{2}\right) \rho
_{0}}\left[ H_{0}\left( \frac{\sqrt{r^{2}+D^{2}}}{\rho _{0}}\right)
-Y_{0}\left( \frac{\sqrt{r^{2}+D^{2}}}{\rho _{0}}\right) \right] \text{ \ or
\ }V(r)=-\frac{ke^{2}}{\varepsilon \sqrt{r^{2}+D^{2}}}.  \label{KeldyshD}
\end{equation}%
for the RK or Coulomb electron$-$hole interaction, respectively. Recently
electrostatic interactions in a bilayer system of TMDC material, which is a
generalization of the RK potential was suggested in Ref.  \cite{Hunt2018}.

It is reasonable in a double layer system when the separation between layers
is big enough to consider the oscillatory approximation (OA) for the RK, as
well as the Coulomb potentials \cite{BermanKezerashviliFBS2011,
RKPhysRevB2012, RKPhysRevB2016, RKPhysRevB2017}.
%in a double layer system when the separation between layer is big enough.
Assuming that $r\ll D$, one can expand Eq.\ (\ref{KeldyshD}) as a Taylor
series in terms of $\left( r/D\right) ^{2}$. By limiting ourselves to the
first order with respect to $\left( r/D\right) ^{2}$, we obtain

\begin{equation}
V(r)=-V_{0}+\beta r^{2},  \label{OCsillator}
\end{equation}%
where

\begin{eqnarray}
V_{0} &=&\frac{\pi ke^{2}}{\left( \varepsilon _{1}+\varepsilon _{2}\right)
\rho _{0}}\left[ H_{0}\left( \frac{D}{\rho _{0}}\right) -Y_{0}\left( \frac{D%
}{\rho _{0}}\right) \right] ,\text{ }\beta =-\frac{\pi ke^{2}}{2\left(
\varepsilon _{1}+\varepsilon _{2}\right) \rho _{0}^{2}D}\left[ H_{-1}\left(
\frac{D}{\rho _{0}}\right) -Y_{-1}\left( \frac{D}{\rho _{0}}\right) \right] ;
\label{V0_C} \\
V_{0} &=&\frac{ke^{2}}{\varepsilon D},\text{ \ }\beta =\frac{ke^{2}}{%
2\varepsilon D^{3}}.  \label{V0_RK}
\end{eqnarray}%
Eqs.~(\ref{V0_C}) and (\ref{V0_RK}) define parameters $V_{0}$ and $\beta $
for the oscillatory approximation of the RK and Coulomb potentials,
respectively, and $H_{-1}\left( \frac{D}{\rho _{0}}\right) $ and $%
Y_{-1}\left( \frac{D}{\rho _{0}}\right) $ are Struve and Bessel functions of
the second kind of order $\nu =-1$, correspondingly.

\begin{figure}[t]
\centering
%\subfigure[]{
\includegraphics[width=6.5 cm]{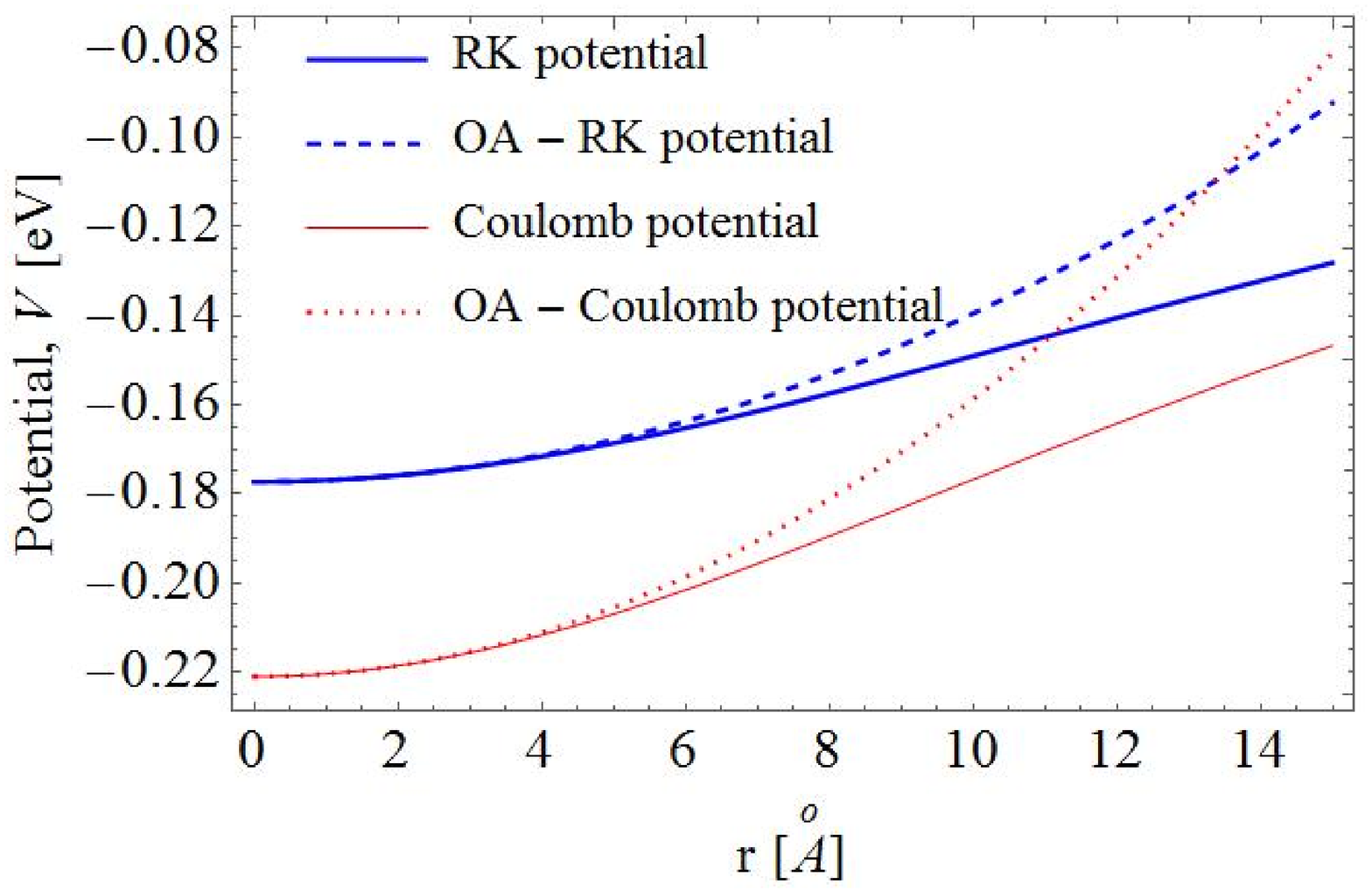} %} \subfigure[]{
\includegraphics[width=6.5 cm]{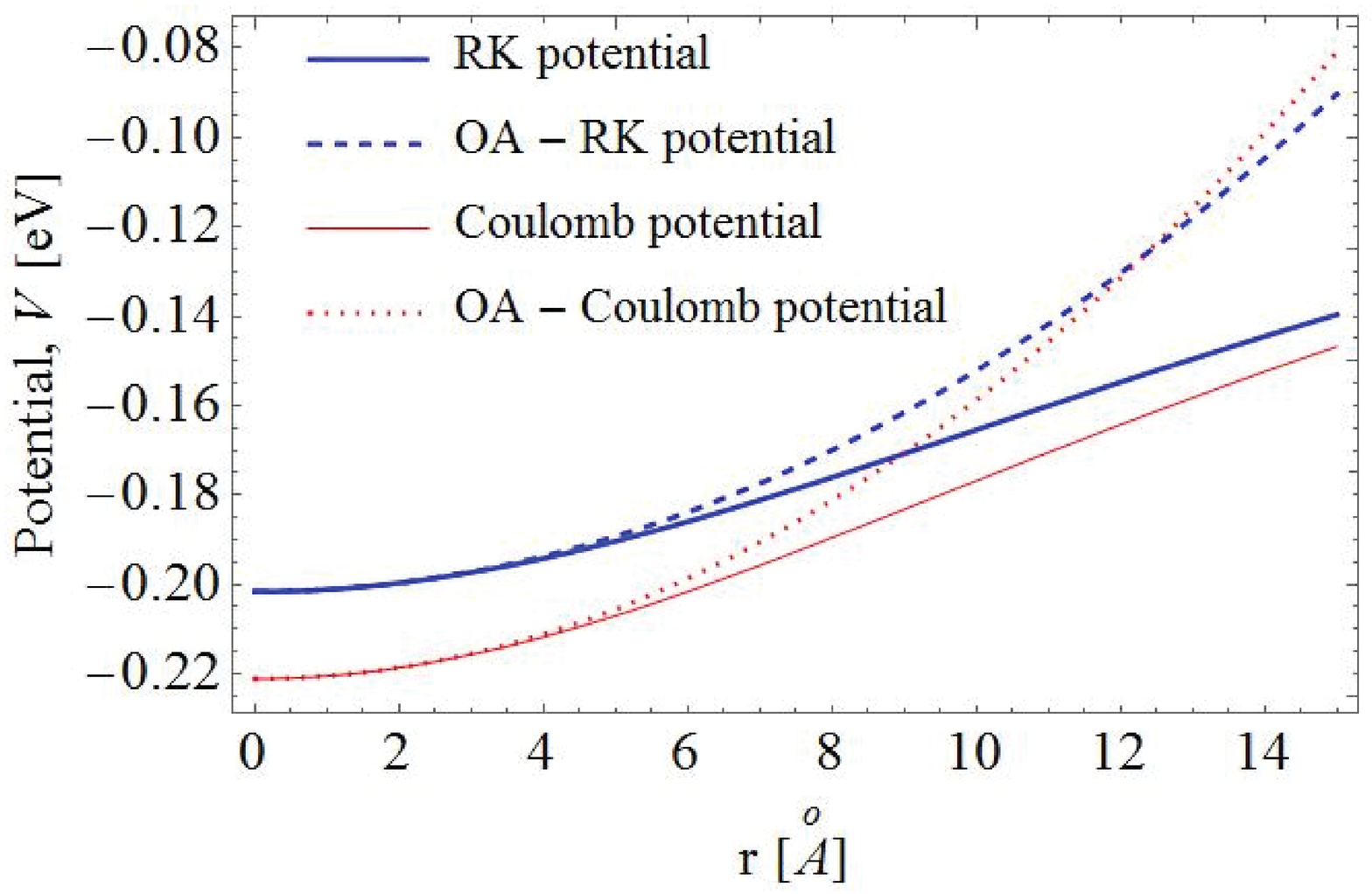} %}
\caption{ The comparison of the RK and Coulomb potentials and their
oscillatory approximations for the electron-hole interaction in MoSe$_{2}$
(a) and phosphorene (b) double layer. At $r < 6$ \AA\ the oscillatory
approximation (3) is a good approximation for both the RK and Coulomb
potentials. At $r > 10$ \AA\ the RK and Coulomb potentials converge each
other as $r$ increases.}
\label{Fig_RK_C}
\end{figure}

Comparisons of the RK and Coulomb potentials for an electron-hole pair in a
MoSe$_{2}$ and phosphorene double layer, as well as their oscillatory
approximation are shown in Fig.~\ref{Fig_RK_C}. According to Fig.~\ref%
{Fig_RK_C}, the RK potential is weaker than the Coulomb potential at small
projections $r$ of the electron-hole distance on the monolayer plane, while
both potentials converge to each other as $r$ increases. The electron-hole
attraction potentials for a MoSe$_{2}$ double layer are significantly larger
than for a phosphorene double layer. The difference between the
Rytova-Keldysh and Coulomb potentials for the phosphorene double layer is
much larger than for the MoSe$_{2}$ double layer, while the Coulomb
potentials for both materials must be the same. In Fig.~\ref{Fig_RK_C} is
considered the case when two monolayers are separated by 4 h-BN monolayers,
corresponding to $D=1.332\ \mathrm{nm}$.

In the low-energy limit, low-energy excitations (e.g. electrons and holes)
in gapless graphene electrons and holes behave as relativistic massless
particles described by the Weyl equation for massless and chiral particles
\cite{Guinea, Vozmediano}, while in gapped 2D materials, excitons are
described by a Dirac-like equation \cite{Kormanyos, Xiao, Tabert}.
Therefore, the physics around the K and K$^{\prime }$ points has attracted
the most attention both experimentally and theoretically. The two-band
single electron Hamiltonian in the $k\cdot p$ approximation in the vicinity
of the K/K$^{\prime }$ points was introduced in Ref. \cite{Xiao} for TMDCs
and Ref. \cite{Tabert} for a buckled honeycomb lattice in the presence of a
perpendicular electric field and are given respectively as

\begin{eqnarray}
H &=&at\left( \xi k_{x}\hat{\sigma}_{x}+k_{y}\hat{\sigma}_{y}\right) +\frac{%
\Delta }{2}\hat{\sigma}_{z}-\lambda \xi \frac{\hat{\sigma}_{z}-1}{2}\hat{s}%
_{z},  \label{TMDC} \\
H &=&\hbar v_{F}(\xi k_{x}\hat{\tau}_{x}+k_{y}\hat{\tau}_{y})-\xi \Delta
_{so}\hat{\sigma}_{z}\tau _{z}+\Delta _{z}\tau _{z}.  \label{Xenes}
\end{eqnarray}%
In Eqs.~(\ref{TMDC}) and (\ref{Xenes}) $\hat{\tau}$ and $\hat{\sigma}$ are
the Pauli matrices for the spin and pseudospin, respectively, $k_{x}$ and $%
k_{y}$ are the components of momentum in the $xy$-plane of the monolayer,
relative to the K and K$^{\prime }$ points, $\xi =1$ ($\xi =-1)$ is the
valley index and denotes the valley K (K$^{\prime }$), $2\lambda $ is the
spin splitting at the valence band caused by spin-orbit coupling in
TMDCs, while in the Xenes, $2\Delta _{so}$ is the intrinsic gap between the
conduction and valence bands at zero electric field, and is also the
splitting due to spin-orbit coupling between the two conduction or two
valence bands at large electric fields. Let us also mention that in (\ref%
{TMDC}) $a$ is the lattice constant, $t$ is the effective hopping integral
(these parameters for a set of the most common layered transition metal
dichalcogenides $\mathrm{MoS_{2}}$, $\mathrm{MoSe_{2}}$, $\mathrm{MoTe_{2},}$
$\mathrm{WS_{2}}$, $\mathrm{WSe_{2}}$ and $\mathrm{WTe_{2}}$ are listed in
Refs. \cite{Kormanyos, Xiao}), $\Delta $ is the energy gap, $\hat{s}_{z}$ is
the Pauli matrix for spin that remains a good quantum number, while in Eq. (%
\ref{Xenes}) $\Delta _{z}=ed_{0}E_{\perp }$ is the gap induced by the
electric field, $E_{\perp },$ normal to a monolayer, where $d_{0}$ in the
latter expression is the buckling constant \cite{Tabert}. The first term in
Eq.~(\ref{Xenes}) is the same as that of the low-energy Hamiltonian in
graphene~\cite{Guinea, Vozmediano}. The second term in (\ref{TMDC}) present
the energy gap in TMDC, while for the Xenes in~(\ref{Xenes}) the second term
describes the intrinsic band gap and the third term gives the modification
of the band gap due to the external electric field. One should emphasize
that in Eq. (\ref{Xenes}) as well as in the Weyl's type equation for the
gapless graphene $v_{F}$ is the Fermi velocity, in contrast to speed of
light in the Weyl and Dirac equations. Therefore, the Hamiltonians (\ref%
{TMDC}) and (\ref{Xenes}), as well as the Hamiltonians for the gapped and
gapless graphene are not relativistically invariant.

Excitons in monolayers and in heterostructures of these monolayers are
usually treated using one of two theoretical methods: the standard quantum
mechanical approach, where the Schr\"{o}dinger equation for an interacting
electron and hole is solved in the framework of the effective mass
approximation, and as a "quasirelativistic" system of two coupled Dirac
particles. The first approach, involving separation of the center-of-mass
and relative coordinates, with a scalar interparticle potential, is
completely understood and is well-developed in configuration and momentum
space in 3D as well as 2D \cite{Landau, Liboff, Griffiths}. By contrast, the
description of the relativistic two-body problem is much more complicated
and until now no completely self-consistent formalism for the separation of
center-of-mass and relative coordinates has been developed, even for the
two-body case. This is due to the following facts stemming from the
relativistic treatment of the electron and hole \cite{RKPhysRevA2012}:

i) the particles' locations and momenta are 4-vectors;

ii) the momenta are not independent but must satisfy mass-shell conditions;

iii) the inter-particle interaction potentials appear in the boosts as well
as in the energy generator in the instant form of dynamics. As a result of a
transformation to the center-of-mass system, even a scalar inter-particle
potential becomes dependent on both a coordinate and momentum;

iv) the structure of the Poincare' group implies that there is no definition
of relativistic 4-center of mass sharing all the properties of the
non-relativistic 3-center of mass \cite{Alba}.

The two-body problem in condensed matter physics for the electron and hole
with the Hamiltonians (\ref{TMDC}) and (\ref{Xenes}) becomes even more
complicated than in the simple relativistic case. It is related to the
following: i) resultant equations from the Hamiltonians (\ref{TMDC}) and (%
\ref{Xenes}) become non-covariant and canonical transformation implementing
the separation of the center-of-mass from the relative variables within
relativistic approach is invalidated; ii) even though the electron-hole
interaction in the RK or Coulomb potentials depends only on the coordinate
of the relative motion, after the center-of-mass transformation the
potential acquires a dependence on the momenta and also due to the chiral
nature of charge carriers one cannot separate the center-of-mass and
relative motions.

\bigskip Based on the single-particle Hamiltonians (\ref{TMDC}) and (\ref%
{Xenes}) one can write the Hamiltonian for the interacting electron-hole
system. The general form of this Hamiltonian\ in the vicinity of the $%
K/K^{\prime }$ points for direct or indirect excitons formed by spin-up
(spin-down) particles is the following:

\begin{equation}
\mathcal{H}_{\uparrow (\downarrow )}=\left(
\begin{array}{cccc}
V(r) & \partial _{2} & \partial _{1} & 0 \\
\partial _{2}^{\dagger } & -\Delta ^{\prime }+V(r) & 0 & \partial _{1} \\
\partial _{1}^{\dagger } & 0 & \Delta ^{\prime }+V(r) & \partial _{2} \\
0 & \partial _{1}^{\dagger } & \partial _{2}^{\dagger } & V(r)%
\end{array}%
\right) \ ,  \label{Rk20}
\end{equation}%
where $V(r)$ is the potential energy of the attraction between an electron
and a hole, which is given by Eq. (\ref{RKeldysh}) or by the Coulomb
potential in the case of direct excitons or by Eqs. (\ref{KeldyshD})
in the case of indirect excitons. In Eq. (\ref{Rk20}) the parameter $\Delta
^{\prime }$ is defined as $\Delta ^{\prime }=\Delta -\lambda $ ($\Delta
^{\prime }=-2\Delta _{z}+2\Delta _{so})$ for spin-up particles, and $\Delta
^{\prime }=\Delta +\lambda $ ($\Delta ^{\prime }=2\Delta _{z}-2\Delta _{so})$
for spin-down particles for TMDCs (Xenes). In Eq.~(\ref{Rk20}) $\partial
_{1}=C(-i\partial _{x_{1}}-\partial _{y_{1}})$, $\partial _{2}=C(-i\partial
_{x_{2}}-\partial _{y_{2}})$ and the corresponding Hermitian conjugates are $%
\partial _{1}^{\dagger }=C(-i\partial _{x_{1}}+\partial _{y_{1}})$, $%
\partial _{2}^{\dagger }=C(-i\partial _{x_{2}}+\partial _{y_{2}})$, where $C$
is a constant and $C=at$ for the TMDC and $C=\hbar v_{F}$ for the Xenes and
graphene. Operators are defined as $\partial _{x}=\partial /\partial x$ and $%
\partial _{y}=\partial /\partial y,$ when $x_{1}$, $y_{1}$ and $x_{2}$, $%
y_{2}$ are the coordinates of vectors $\mathbf{r}_{1}$ and $\mathbf{r}_{2}$
for an electron and hole, correspondingly.

The Hamiltonian (\ref{Rk20}) describes two interacting particles located in
2D monolayers or 2D double layers and satisfies the following conditions:

i) when the potential $V(r)=0$, the Hamiltonian describes two
non-interacting Dirac particles in mono or double layers.

ii) when $\Delta ^{\prime }=0$ and $V(r)$ is the Coulomb potential, the
Hamiltonian describes two interacting Dirac particles in gapless graphene
and is identical to the Hamiltonian~\cite{Sabio} representing the two-body
problem in gapless graphene layer;

iii) depending on the values of $\Delta ^{\prime }$ the Hamiltonian (\ref%
{Rk20}) describes the interacting the electron-hole system via the RK
potential \cite{Keldysh} in a monolayer TMDC or Xenes. In the case of
indirect excitons, when the electron and hole are located in two different
monolayers with the interlayer separation $D$, one can consider the
electron-hole interaction via the RK\ or Coulomb potentials (\ref{KeldyshD})
or use the OA (\ref{OCsillator}).

The energy spectrum of an electron-hole pair can be found by solving the
eigenvalue problem for the Hamiltonian~(\ref{Rk20}):

\begin{equation}
\mathcal{H}_{\uparrow (\downarrow )}\Psi _{\uparrow (\downarrow )}=\epsilon
_{\uparrow (\downarrow )}\Psi _{\uparrow (\downarrow )}\ ,  \label{Shr1}
\end{equation}%
where $\epsilon _{\uparrow (\downarrow )}$ is the energy spectrum for an
electron-hole pair with the up and down spin orientation. The eigenfunction $%
\Psi _{\uparrow (\downarrow )}$ in Eq. (\ref{Rk20}) is four-component
spinor, where the spinor components refer to the four possible values of the
conduction/valence band indices and is given as:
\begin{equation}
\Psi _{\uparrow }(\mathbf{r}_{1},\mathbf{r}_{2})=\left( {%
\begin{array}{c}
\phi _{c\uparrow c\uparrow }(\mathbf{r}_{1},\mathbf{r}_{2}) \\
\phi _{c\uparrow v\uparrow }(\mathbf{r}_{1},\mathbf{r}_{2}) \\
\phi _{v\uparrow c\uparrow }(\mathbf{r}_{1},\mathbf{r}_{2}) \\
\phi _{v\uparrow v\uparrow }(\mathbf{r}_{1},\mathbf{r}_{2})%
\end{array}%
}\right) \equiv \left( {%
\begin{array}{c}
\Psi _{c\uparrow } \\
\Psi _{v\uparrow }%
\end{array}%
}\right) ,\text{ where }\Psi _{c\uparrow }=\left( {%
\begin{array}{c}
\phi _{c\uparrow c\uparrow } \\
\phi _{c\uparrow v\uparrow }%
\end{array}%
}\right) ,\ \ \ \Psi _{v\uparrow }=\left( {%
\begin{array}{c}
\phi _{v\uparrow c\uparrow } \\
\phi _{v\uparrow v\uparrow }%
\end{array}%
}\right) ,  \label{wave function1}
\end{equation}%
where a quasiparticle is characterized by the coordinates $\mathbf{r}_{j}$
in the conduction ($c$) and valence ($v$) band with the corresponding
direction of spin up $\uparrow $ or down $\downarrow $, and index $j=1,2$
referring to the two monolayers, one with electrons and the other with
holes. In this notation we assume that a spin-up (-down) hole describes the
absence of a spin-down (-up) valence electron. The two components reflect
one particle being in the conduction (valence) band and the other particle
being in the valence (conduction) band, correspondingly. Let us mention that
while (\ref{wave function1}) represents the spin-up particles, the
spin-down particles are represented by the same expression replacing $%
\uparrow $ by $\downarrow $.

As aforementioned, for the Hamiltonian~(\ref{Rk20}) the center-of-mass
motion cannot be separated from the relative motion in the quasirelativistic
approach due the chiral nature of charge carriers in 2D materials. A similar
conclusion was made for the two-particle problem in graphene ~\cite{Sabio},
gapped graphene \cite{RKPhysRevB2012} and TMDC monolayers \cite%
{RKPhysRevB2016}. Since the RK and Coulomb interactions depend only on the
relative coordinate of the electron-hole system, one can introduce the
\textquotedblleft center-of-mass\textquotedblright\ coordinate and the
relative motion coordinate in the plane of a monolayer: $\mathbf{R}=\alpha
\mathbf{r}_{1}+\delta \mathbf{r}_{2},\mathbf{r}=\mathbf{r}_{1}-\mathbf{r}%
_{2} $, where the coefficients $\alpha $ and $\delta $ are supposed to be
found for a small momentum $\mathbf{K}$ from the condition of the separation
of the coordinates of the center-of-mass and relative motion of an
electron-hole in the one-dimensional equation for the corresponding
components of the spinor $\Psi _{\uparrow (\downarrow )}$ \cite%
{RKPhysRevA2012}. One can obtain the solution of Eq.~(\ref{Shr1}) by making
the following Anz\"{a}tze
%%%%%%%%%%%%%%%%%%%%%%%%%%%%%%%%%%%%%%%%%%%%%%%%%%%%%%%%%%%%%%%%%%%%%%%%%%%%%%%%%%%%%%%%%%%%%%%%%%%

\begin{equation*}
\Psi _{j\uparrow (\downarrow )}(\mathbf{R},\mathbf{r})=\mathtt{e}^{i\mathbf{K%
}\cdot \mathbf{R}}\psi _{j\uparrow (\downarrow )}(\mathbf{r})\ ,
\end{equation*}%
and follow the procedure given in Refs. \cite{RKPhysRevA2012,
RKPhysRevB2012, RKPhysRevB2016} one derives an equation for the components $%
\phi _{c\uparrow (\downarrow )v\uparrow (\downarrow )}$ of the spinor $\Psi
_{\uparrow (\downarrow )}.$ As an example, let us introduce the equation for
the components $\phi _{c\uparrow (\downarrow )v\uparrow (\downarrow )}$ of
the bound electron-hole system for TMDC materials \cite{RKPhysRevB2016}:

\begin{equation}
\left( -F_{1}(\epsilon _{\uparrow (\downarrow )})\nabla _{\mathbf{r}%
}^{2}+V(r)\right) \phi _{c\uparrow (\downarrow )v\uparrow (\downarrow
)}=F_{0}^{\prime }(\epsilon _{\uparrow (\downarrow )})\phi _{c\uparrow
(\downarrow )v\uparrow (\downarrow )}\ ,  \label{R26}
\end{equation}%
where

\begin{equation}
F_{1}(\epsilon _{\uparrow (\downarrow )})=\frac{2a^{2}t^{2}}{\epsilon
_{\uparrow (\downarrow )}}\ ,\text{ \ \ \ \ \ }F_{0}^{\prime }(\epsilon
_{\uparrow (\downarrow )})=\epsilon _{\uparrow (\downarrow )}+\Delta
^{\prime }-\frac{a^{2}t^{2}\mathcal{K}^{2}}{2\epsilon _{\uparrow (\downarrow
)}}\ .
\end{equation}

\subsubsection{\protect\bigskip Double layers of TMDC}

For the two-body electron-hole system interacting via the Rytova-Keldysh
potential, Eq. (\ref{R26}) has no analytical solution and can only be solved
numerically, while for the Coulomb potential one can obtain an analytical
solution \cite{RKPhysRevA2012}. For indirect excitons, (\ref{R26}) can be
solved only numerically for both types of potentials. One can consider a
spatially separated electron-hole pair in two parallel TMDC layers at large
distances $D\gg a_{B}$, where $a_{B}$ is the 2D Bohr radius of a dipolar
exciton and use the oscillatory approximation (\ref{OCsillator}). For TMDC
materials the Bohr radius of the dipolar exciton is found to be in the range
from $1.5\ \mathrm{{\mathring{A}}}$ for $\mathrm{MoTe_{2}}$~\cite{Ashwin} up
to $3.9\ \mathrm{{\mathring{A}}}$ for $\mathrm{MoS_{2}}$~\cite{Louie}.
Therefore, one can use the OA (\ref{OCsillator}), which allows one to reduce
the problem of indirect exciton to an exactly solvable two-body problem. By
substituting~(\ref{OCsillator}) into Eq.~(\ref{R26}), one obtains an
equation that has the form of the Schr\"{o}dinger equation for the 2D
isotropic harmonic oscillator:
%%%%%%%%%%%%%%%%%%%%%%%%%%%%%%%%%%%%%%%%%%%%%%%%%%%%%%%%%%%%%%%%%%%%%%%%%%%%%%%%%%%%%%%%%%%%%%%%%%%%%

\begin{equation}
\left( -F_{1}(\epsilon _{\uparrow (\downarrow )})\nabla _{\mathbf{r}%
}^{2}+\beta r^{2}\right) \phi _{c\uparrow (\downarrow )v\uparrow (\downarrow
)}=F_{0}(\epsilon _{\uparrow (\downarrow )})\phi _{c\uparrow (\downarrow
)v\uparrow (\downarrow )}\ ,
\end{equation}%
where $F_{0}(\epsilon _{\uparrow (\downarrow )})=F_{0}^{\prime }(\epsilon
_{\uparrow (\downarrow )})+V_{0},$ and parameters $V_{0}$ and $\beta $ are
given by Eqs. (\ref{V0_C}) and (\ref{V0_RK}) for \ both the Rytova-Keldysh
and Coulomb potentials, respectively.

The solution of the Schr\"{o}dinger equation for the harmonic oscillator, is
well known and is given by
%%%%%%%%%%%%%%%%%%%%%%%%%%%%%%%%%%%%%%%%%%%%%%%%%%%%%%%%%%%%%%%%%%%%%%%%%%%%%%%%%%%%%%%%%%%%%%%%%%%%%%%%%%%%%%%%%%%%%%%%%%%%%%%%%%%%%%%%%%%%%%%%%%%%%%%%%%
\begin{equation}
\frac{F_{0}(\epsilon _{\uparrow (\downarrow )})}{F_{1}(\epsilon _{\uparrow
(\downarrow )})}=2N\sqrt{\frac{\beta }{F_{1}(\epsilon _{\uparrow (\downarrow
)})}}\ ,
\end{equation}%
where $N=2\tilde{N}+|L|+1$, and $\tilde{N}=\mathrm{min}(\widetilde{n},%
\widetilde{n}^{\prime })$, $L=\widetilde{n}-\widetilde{n}^{\prime }$, $%
\widetilde{n},$ $\widetilde{n}^{\prime }=0,1,2,3,\ldots $ are the quantum
numbers of the 2D harmonic oscillator. The corresponding 2D wave function
can be expressed in terms of associated Laguerre polynomials \cite%
{RKPhysRevB2016}. \ Thus, considering Eq. (\ref{R26}) for indirect excitons
and using the oscillatory approximation, we can reduce the problem of
indirect exciton to an exactly solvable two-body problem. The binding energy
for the indirect exciton was estimated for two $\mathrm{MoS_{2}}$ layers
separated by $N$ h-BN insulating layers from $N=1$ up to $N=6$~\cite{Fogler}%
. These dipolar excitons were observed experimentally for $N=2$~\cite{Calman}%
. We assume that the indirect excitons in TMDCs can survive for a larger
interlayer separation $D$ than in semiconductor coupled quantum wells,
because the thickness of a TMDC layer is fixed, while the spatial
fluctuations of the thickness of the semiconductor quantum well affects the
stability of the dipolar exciton. The theoretical analysis presented above
is quite general and can be applied to any TMDC monolayer or a double layer
system with two different Mo- and W-based monolayers. The effect of
different dielectric environments on the exciton binding energy in the
framework of a four-band Hamiltonian describing indirect excitons is
investigated and a remarkable dependence on the dielectric constant of the
barrier between the two layers is found \cite{Peeters2018}.

The description of the Mott-Wannier excitons within the effective mass
potential model requires two main inputs of material-specific ingredients
such as electrons and holes effective masses, which are easily calculated
from \textit{ab initio} band structures, and polarizability of the
monolayer, which is an essential parameter for the description of the
screened electron-hole interaction. The screening effects are negligible for
electron-hole distances larger than the screening length $\rho _{0}$, and at
long range the electron-hole interaction is described by the Coulomb
potential~\cite{Reichman2013}. The screening length is defined by the 2D
polarizability of the planar material~\cite{Rubio}. Using the
polarizabilities from Ref.~\cite{Reichman2013}, we conclude that $\rho _{0}$
is estimated as $38\ \mathrm{{\mathring{A}}}$ for $\mathrm{WS_{2}}$, $41\
\mathrm{{\mathring{A}}}$ for $\mathrm{MoS_{2}}$, $45\ \mathrm{{\mathring{A}}}
$ for $\mathrm{WSe_{2}}$, $52\ \mathrm{{\mathring{A}}}$ for $\mathrm{MoSe_{2}%
}$. The spin-orbit coupling in TMDC monolayers leads to a spin-orbit
splitting in the valence band and to the formation of two distinct types $A$
and $B$ excitons \cite{Reichman2013, Kormanyos}. $A$ excitons are formed by
spin-up electrons from conduction and spin-down holes from valence band,
while type $B$ excitons are formed by spin-down electrons from conduction
and spin-up holes from valence band \cite{GlazovRMP}. Comparing the binding
energies of excitons over different TMDC layers, one can observe that the
binding energy depends weakly on the effective masses, but strongly on the
material polarizability. As a consequence, the binding energy for $A$ and $B$
excitons, which have different effective masses are generally very similar
in the same monolayer \cite{DenFuncTheoryPIMC, BrunettiJP2018}.

\subsubsection{Phosphorene}

The exciton binding energy of monolayer phosphorene on a SiO$_{2}$/Si
substrate was determined to be $\sim $0.9 eV \cite{WangPhosphorenEx}. This
result agrees well with the theoretical prediction that substrate screening
strongly affects the exciton binding energy in monolayer phosphorene \cite%
{Rodin}. A theoretical study of the exciton binding energy using the
screened electron-hole interaction for anisotropic two-dimensional crystals
is presented in Ref. \cite{Prada}, where the authors obtained analytical
expressions using variational wave functions in different limits of the
screening length. The analytical solution for the exciton binding energy
using the variational approach \cite{Prada} gives a result which compares
well with the numerical one and is in reasonable agreement with the
experimental value for the monolayer of black phosphorous. A recent \textit{%
ab initio} study \cite{Hunt22019} has used the diffusion Monte Carlo method
to study the exciton binding energy of monolayer phosphorene from
first-principles. The double layer phosphorene system with a number of h-BN
monolayers, placed between two phosphorene monolayers was investigated in
Ref. \cite{RKPhysRevBG2017}. The different lattice constants from the
literature in turn causes the difference in the band curvatures, and,
therefore, in the different effective masses of charge carriers. The binding
energies of indirect excitons formed in the double layer phosphorene with 7
h-BN monolayers, calculated for the sets of the different masses
corresponding to the different lattice constants\ are 28.2 meV, 29.6 meV,
37.6 meV, and 37.2 meV \cite{RKPhysRevBG2017}.

\subsubsection{Xenes}

Related to the 2D Xenes, the study of binding energies and optical
properties of direct and indirect excitons in monolayers and double layer
heterostructures of Xenes (silicene, germanene, and stanene) is presented in
Ref. \cite{Brunetti2018},\ where the Schr\"{o}dinger equation with electric
field-dependent exciton reduced mass is solved by using the RK potential for
direct excitons, while both the RK and Coulomb potentials are used for
indirect excitons. \ One of the important features of the 2D semiconductors
is the existence of strongly bound excitons with binding energies reaching
up to 30\% of the band gap. However, calculations of the binding energies
with the RK potential as a function of external electric field in
freestanding Xenes \cite{Brunetti2019} \ demonstrate that these binding
energies are far larger than their respective band gaps, when the electric
field is small or zero. This phenomena could be an indicator of the
excitonic insulating phase in these materials. Thus, one can observe a phase
transition in monolayer Xenes from the excitonic insulator ground state to
the semiconducting phase by increasing the electric field beyond some
critical value which is unique to each material. In the case of an Xene
monolayer on a substrate, the enhanced dielectric screening from the
substrate reduces the exciton binding energy such that Xenes on a substrate
should not exhibit the excitonic insulator phase.

\subsection{1D excitons}

To model the electrostatic interaction of an electron and hole in 1D quantum
system, Poisson's equation is first solved to find the electrostatic
potential of a point-like charge $e$ inside a 1D structure, which is
considered as a dielectric cylinder. The cumbersome nature of such a
potential makes it impossible to solve two-body problem analytically.
However, it can be solved when the complicated potential is approximated by
an effective potential. One first calculates the 1D subband energies and
wave functions, while neglecting the Coulomb interaction, and using these
wave functions of transverse electron and hole motion, calculates the
longitudinal motion of the exciton, including corrections from image forces
in the surrounding medium. To do that, the three dimensional Coulomb
potential is averaged to a one dimensional Coulomb interaction between the
electron and hole along the 1D nanostructure axis \cite{Bartnik, Giblin}.
Thus, to perform the calculations, the Coulomb potential is often replaced
by approximate model potentials. There are different models of the effective
interaction potential: i. The effective 1D electron-hole interaction is
modeled as cusp-type Coulomb potential $V(z)=-\frac{ke^{2}}{4\pi \varepsilon
_{0}\varepsilon }\frac{A}{z-r_{0}}$, where the parameters $A$ and $r_{0}$
are determined self-consistently by employing the eigenfunctions of the
lateral confinement of electrons and holes \cite{OgawaTakagahara, RK1D}; ii.
The singularity of the Coulomb potential is cut off at $r=a$, where $a$ is
the radius of the wire, and the effective potential is $V(z)=-\frac{ke^{2}}{%
4\pi \varepsilon _{0}\varepsilon }\frac{1}{\sqrt{z^{2}+a^{2}}}$ \cite%
{DasSarma, Semina}; iii. An assumption of strong lateral confinement, allows
separation of the $z$ motion from the lateral motion in the $xy$ plane, and
by averaging the 3D Coulomb interaction potential over the transverse
degrees of freedom, an analytical 1D formula for the effective interaction
potential between the confined charge carriers is derived \cite%
{Bednarek2003, Slachmuylders}. iv. The potential is divided into four terms:
the unscreened direct interaction of the two charges, the modification of
this interaction due to the image effects, and the two self-interactions of
each charge with its own image, and the adiabatic potential is obtained by
averaging the potential over wave functions of the corresponding electron
and hole subband \cite{Bartnik}. However, in all cases the electromagnetic
interaction between the charge carriers is not given by the 3D Coulomb
potential but rather by one-dimensional potentials, which have a Coulomb
tail, and consequently the highly excited bound eigenstates of 1D excitons
have a Rydberg series structure. Calculations show that for charge carriers
confined in 1D nanostructures, the aforementioned effective 1D model
potentials work with a reasonable precision in a wide range of nanostructure
parameters. Fig. 5 depicts the electric field lines between the interacting
electron and hole. For 2D materials, field lines are screened along the plane
and mainly lie unscreened in the vacuum. In 1D materials, field lines lie
mainly in the vacuum, hence screening is heavily suppressed.

\begin{figure}[t]
\centering
\par
\includegraphics[width=0.30\textwidth]{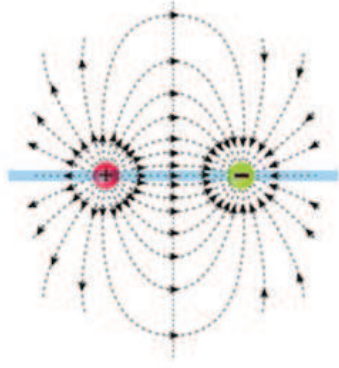}
\caption{ Electric field lines for two interacting particles in 1D
materials.}
\label{F5}
\end{figure}

Following \cite{OgawaTakagahara} one can write the equation that describes
one-dimensional relative motion of the electron and hole interacting via a
cusp-type Coulomb interaction as

\begin{equation}
-\frac{\hbar ^{2}}{2\mu }\frac{d^{2}\Phi _{X}(z)}{dz^{2}}-\frac{A_{eh}}{%
\left\vert z\right\vert +Z_{0eh}}\Phi _{X}(z)=E_{X}\Phi _{X}(z).
\label{Exciton}
\end{equation}%
Here $\mu $ is the reduced effective mass of the electron-hole pair, $%
A_{eh}, $ $Z_{0eh}$ are the fitting parameters for the effective electron$-$%
hole one-dimensional cusp-type Coulomb potentials obtained through the
parametrization, $z=z_{e}-z_{h}$ is the relative electron-hole motion
coordinate, $-E_{X}$ is the binding energy of the exciton, and $\Phi _{X}(z)$
is the corresponding excitonic eigenfunction. Eq. (\ref{Exciton}) has the
same form as the equation for one-dimensional hydrogen atom studied by
Loudon \cite{Loudon1}. One can introduce the following notations

\begin{equation}
\xi ^{2}=-\frac{\hbar ^{2}\eta _{0}^{2}}{2\mu E_{X}},\text{ }\eta _{0}=\frac{%
A_{eh}\mu }{\hbar ^{2}},\text{ }x=\frac{2\eta _{0}(\left\vert z\right\vert
+Z_{eh})}{\xi }  \label{Notation}
\end{equation}%
and reduce (\ref{Exciton}) to the Whittaker's equation

\begin{equation}
\frac{d^{2}\zeta (x)}{dx^{2}}+\left( -\frac{1}{4}+\frac{\xi }{x}\right)
\zeta (x)=0.  \label{Whittaker}
\end{equation}%
The solution of (\ref{Whittaker}) is $\zeta (x)=W_{\xi ,\pm 1/2}(x)$ \cite%
{Loudon1, Loudon2, Loudon3}, where $W_{\xi ,\pm 1/2}(x)$ is the Whittaker
function. The value of $\xi $ which defines $E_{X}$\ and $\Phi _{X}(z)$, is
determined by the boundary condition stating that for even states the
derivative of wavefunction at $z=0$ must turn to zero

\begin{equation}
\frac{d}{dz}\left[ W_{\xi ,\pm 1/2}\left( \frac{2\eta _{0}(\left\vert
z\right\vert +Z_{eh})}{\xi }\right) \right] _{z=0}=0.  \label{ExcitEner}
\end{equation}

The full Hamiltonian of excitons in a quantum nanowires (NWR) is constructed
within $k\cdot p$ theory using the single-band effective mass approximation
or even four-band effective mass model \cite{Bartnik}. The formation,
stability, and binding energy of excitons depends on the electron to hole
mass ratio and the geometric characteristics of a nanostructure, such as the
shape of the NWR, the radius of the NWR or carbon nanotudes (CNT), and the
thickness of a shell for core/shell NWR. The excitons in NWRs and CNTs are
well studied objects and theoretical calculations have shown that the
effective 1D interaction leads to accurate results for different
characteristics of NWRs as well as CNTs, which are in reasonable agreement
with available experimental data.

\section{Three Body Problem$-$Trions}

In the late 1950s Lampert \cite{L} predicted the existence of charged
three-particle complexes: a negatively (X$^{-}$) and positively (X$^{+}$)
charged trions, formed when an electron in a conduction band or a hole in a
valence band is bound to an exciton. This idea gave rise to many
publications in the 60s and the 70s to study trions in bulk materials. The
binding energies of these exciton complexes are very small in bulk at room
temperature, but they are substantially enhanced in structures of reduced
dimensionality. Theoretical calculations performed at the end of the 1980s
\cite{Stebe1987, Stebe1989} predicted that confining particles to a quasi-2D
well increases the binding energy considerably due to the increased overlap
of the electron and hole wavefunctions. The confinement increases (up to
tenfold) of the trion binding energy in quantum well (QW) heterostructures
compared to bulk semiconductors. In 1993, 35 years after their prediction,
negatively charged trions were first observed in CdTe quantum wells \cite%
{QW1}, which stimulated intensive experimental and theoretical studies of
trions. However, typically, trions in QW are localized at cryogenic
temperatures and their binding energies are a few meV.

\subsection{3D trions}

Mott--Wannier trions in 3D semiconductors are intrinsically three-particle
objects, which can be described by the solutions of the three-body Schr\"{o}%
dinger equation after modeling the crystal by effective electron and hole
masses and a dielectric constant. Trions present the system with two
identical particles $eeh$ (X$^{-}$) or $ehh$ (X$^{+}$) and, therefore, one
deals with a three-body system \textit{AAB}(\textit{ABB} with two identical particles.
The Faddeev formalism \cite{Faddeev, MF85} is the most rigorous approach for
investigating a three-body system. The differential Faddeev equations \cite%
{MF85} in configuration space can be written in the form of a system of
second order differential equations and have a simpler form for the case of
two identical particles. Introducing the set of the Jacobi coordinates for
the three particles, separating the motion of the center-of-mass from the
relative motion, one can decompose the total wave function of the system
into the sum of the Faddeev components $U$ and $W$ corresponding to the $%
(AA)B$ and $(AB)B$ types of rearrangements: $\Psi =U+W-\hat{P}W$, where $%
\hat{P}$ is the permutation operator for two identical fermions. The set of
the Faddeev equations for $U$ and $W$ components of the relative motion of
three particles when two of them are identical fermions can be written as
\cite{Filikhin, KezTFSV}:

\begin{eqnarray}
(H_{0}+V_{AA}-E)\ U &=&-V_{AA}(W-\hat{P}W),  \notag \\
(H_{0}+V_{AB}-E)\ W &=&-V_{AB}(U-\hat{P}W).  \label{FadTrion}
\end{eqnarray}%
In Eq. (\ref{FadTrion}) $H_{0}$ is the operator of kinetic energy of the
Hamiltonian taken for corresponding Jacobi coordinates, $V_{AA}$ and $V_{AB}$
are the Coulomb potentials with the dielectric constant related to the
considered material.

In the framework of the aforementioned Faddeev formalism the binding
energies for the 3D trions are calculated in Ref. \cite{FilKezPLA2018} using
as the inputs the Coulomb potential, the known effective masses of electrons and holes
obtained for various bulk materials and the corresponding dielectric
constants $\varepsilon $. The theoretical analysis presented in \cite%
{FilKezPLA2018} is quite general and can be applied to any 3D or 2D
materials.\textit{\ }The results of calculations led to a surprising result:
for the same bulk semiconductors the negatively charged trions are weakly
bound, while the positively charged trions are completely unbound for
experimentally known dielectric constants. At the first look this result
seems strange because the binding energy of trions is determined by the
Coulomb potential between the charge carriers of the same electric charge.
The only difference is related to the different masses of the electron and
the hole. For the same electron and hole masses the binding energies of X$%
^{-}$ and X$^{+}$ are equal. The origin of a discrepancy for the binding
energies was addressed by considering a hypothetical model with the
potential $\alpha V_{AA}$ ($0\leq \alpha \leq 1)$, where the parameter $%
\alpha $\ controls the strength of interaction between identical particles
for both trions and effectively leads to a weaker Coulomb repulsion between
the identical particles and hence an increased trion binding energy. Solving
the Faddeev equations with this hypothetical potential, one can find binding
energies for the X$^{-}$and X$^{+}$ trions and test the sensitivity of their
binding energy to the strength of $\alpha V_{AA}$ by varying the parameter $%
\alpha $. The calculated contour plots of the Faddeev component $U$ and $W$
obtained from Eq. (\ref{FadTrion}) demonstrated that two holes are located
more closely to each other in X$^{+}$ than the two electrons in X$^{-}$.
Therefore, the Coulomb repulsion is stronger for two holes than for two
electrons due to the close localization of the holes to each other. In other
words, the mean inter-hole distances are much shorter than those of
electrons, which makes two holes interact repulsively more intensively in
the X$^{+}$ than two electrons do in the X$^{-}$.

\subsection{2D trions}

Development of high-quality semiconductor quantum wells and highly doped
materials allowed precise experimental studies of X$^{-}$ and X$^{+}$ trions
states using optical measurements \cite{QW1, QW2, QW3, QW4, QW5, QW6, QW7,
Esser1, Esser2, Bracker}. Indeed, QW can be considered as quasi-two
dimensional systems and the reduction of dimensionality is known to enhance
trions binding energies \cite{Stebe1989}. Variational approaches are
commonly used to find the trion ground state energy \cite{Stebe1989,
Varga1999, Varga2000, Ronnow}. Elaborate variational methods, made feasible
by rapid increases in computational power, can calculate the trion ground
state energy to an amazing precision of ten decimal places or more. Within
the method of hyperspherical harmonics, a numerically accurate procedure is
proposed to solve the Schr\"{o}dinger equation for charged excitons in
quasi-two dimensions. Numerical results for negatively charged trion are in
good agreement with those obtained through other computationally intensive
methods \cite{HHRuan1999}. A system of three identical charged particles in
a two-dimensional harmonic well and a transverse magnetic field is treated
in the framework of the Faddeev approach in configuration space \cite%
{Braun2001, Braun2004}. In all these studies the Coulomb potential was
considered between charged particles.

Experimental and theoretical interest in trions in 2D materials has
increased dramatically since 2013, when trions have been observed in 2D MoS$%
_{2}$ monolayer \cite{MoS23Heinz}, and their signature has not just appeared
at low temperature but up to room temperature. The positively, and
negatively, charged trions were observed by different experimental groups in
TMDC monolayers including both: molybdenum - (MoS$_{2}$ and MoSe$_{2})$ and
tungsten - (WS$_{2}$ and WSe$_{2}$) based monolayers \cite{MoS23Heinz,
MoSe21 Ross, WSe2 Jones, WSe2Wang, MoSe2Singh, Liu, ShangBiexiton,
WS2Plechinger, ZhangMS2, Christopher2017, CourtadeSemina}. Trions in
monolayer TMDCs are stable at room temperature due to their remarkably large
binding energies in the range of a few tens of meV. In MoS$_{2}$ monolayer a
trion is formed by an exciton with an extra electron or hole, which can be
introduced by gate-doping, photoionization of impurities, or choice of
substrates \cite{Bellus}. These trions have been observed in
photoluminescence, electroluminescence, and absorption spectra. From these
measurements, the trion binding energy in MoS$_{2}$ was found to be in the
range of $18-43$ meV \cite{MoS23Heinz, ZhangMS2, Christopher2017}, while the
binding energy of trions in MoSe$_{2}$ was determined to be 30 meV \cite%
{MoSe21 Ross, MoSe2Singh}. Similarly, trions were also observed in tungsten$-
$based monolayers with binding energies $10-45$ meV \cite{Shang,
WS2Plechinger, bZu} \ and $20-30$ meV \cite{WSe2 Jones, WSe2Wang,
CourtadeSemina}\ \ in WS$_{2}$ and WSe$_{2}$ monolayers, respectively. The
binding energies of X$^{-}$ and X$^{+}$ in monolayer MoTe$_{2}$ were
measured to be $\thicksim 27$ meV and $\thicksim 24$, respectively \cite%
{Yang}. A detailed investigation of the exciton and trion dynamics in MoSe$%
_{2}$ and WSe$_{2}$ monolayers as a function of temperature in the range
10--300 K is presented in Ref. \cite{Godde}. Recently was reported the
observation of excitonic fine structure in a 2D TMDC semiconductors \cite%
{Shang, CourtadeSemina, WS2Plechinger2016}. In photoluminescence and in
energy-dependent Kerr rotation measurements the trion fine structure and
coupled spin--valley dynamics in monolayer WS$_{2}$ was studied \cite%
{WS2Plechinger2016}. This experimental approach was able to resolve two
different trion states, which are interpreted as intravalley and intervalley
trions. A well-resolved fine-structure splitting of 6 meV for the negatively
charged trion in WSe$_{2}$ is recently observed \cite{CourtadeSemina}.

Studies of trions were extended to van der Waals heterostructures formed by
the same TMDC monolayers \cite{bZu} and two different TMDC monolayers.
Tightly bound trions with binding energies of 62 meV in TMDC WS$_{2}/$MoSe$%
_{2}$ heterostructure are formed by excitons excited in the WS$_{2}$ layer
and electrons transferred from the MoSe$_{2}$ layer \cite{Bellus}.

Due to the reduced dimensionality and screening in the highly anisotropic
phosphorene layer, excitons and trions formed in a 2D phosphorene monolayer
exhibit quasi-one-dimensional (quasi-1D) behavior and possess binding
energies that are much larger than those in quasi-2D quantum wells and other
isotropic 2D materials, such as TMDC. A huge trion binding energy of $%
\thicksim $100 meV was first observed in monolayer phosphorene \cite{Yang2
2017}, which is around two to five times higher than that in TMDCs
semiconductors, such as  molybdenum- or tungsten-based monolayers. The
measured ultrahigh trion binding energies in three phosphorene monolayers on
a SiO$_{2}/$Si substrate reported in Ref. \cite{Yang3 2016} is $\thicksim $%
162 meV, which is a result of formation of quasi-1D trions in 2D phosphorene.

The nonrelativisic trion Hamiltonian in a 2D configuration space is given by

\begin{equation}
H=-\frac{\hbar ^{2}}{2}\sum_{i=3}^{3}\frac{1}{m_{i}}\nabla
_{i}^{2}+\sum_{i<j}^{3}V_{ij}(\left\vert \mathbf{r}_{i}-\mathbf{r}%
_{j}\right\vert ),  \label{Trion}
\end{equation}%
where $m_{i}$ is the $i$th charge carrier mass and $\mathbf{r}_{i}$ is the $i
$th particle position in a 2D configuration space. In Eq. (\ref{Trion}) $%
V_{ij}(\left\vert \mathbf{r}_{i}-\mathbf{r}_{j}\right\vert )$ is the
pairwise interaction energy of the charge carriers.  In some theoretical
studies the interaction between charged particles is considered by using the
Coulomb potential, while many researchers use the RK potential \cite{Keldysh}%
, which describes the Coulomb interaction screened by the polarization of
the electron orbitals in the 2D lattice. By introducing the Jacobi
coordinate $\mathbf{x}_{i}$ and $\mathbf{y}_{i}$ in 2D configuration space
and separating the center-of-mass and relative coordinates, the Schr\"{o}%
dinger equation for the relative motion of the three-body system reads

\begin{equation}
\lbrack -\frac{\hbar ^{2}}{2\mu }(\nabla _{x_{i}}^{2}+\nabla
_{y_{i}}^{2})+\sum_{i<j}^{3}V_{ij}(\left\vert \mathbf{r}_{i}-\mathbf{r}%
_{j}\right\vert )-E]\Psi (\mathbf{x}_{i},\mathbf{y}_{j})=0.\ \
\label{Relative3}
\end{equation}%
A variety of theoretical approaches have been proposed for solution of Eq. (%
\ref{Relative3}) and finding the binding energies of trions in 2D materials
by incorporating a proper treatment of screening in two dimensions via RK
potential. Initial work for calculations of trion binding energies in TMDCs
\cite{Reichman2013} was based on an effective mass model and the necessary
parameters for the exciton and trion Hamiltonians were calculated from first
principles. In particular, the trion binding energies were found by means of
a simple few-parameter variational wave functions \cite{Reichman2013,
ZhangExTheor}, \ and later were used variational optimization with more intricate
trial wave functions. An approach to construct the variational trion wave
function in a 2D TMDC semiconductor with requirements imposed by the
symmetry on the permutation of identical particles is presented in Ref. \cite%
{CourtadeSemina} and, using an effective mass model, the authors estimated
trion binding energies of 20 to 30 meV for both X$^{+}$ and X$^{-}$.

The diffusion quantum Monte Carlo approach was used to find the numerical
solution of the Schr\"{o}dinger equation and obtain the energies of
negatively charged trions within the Mott-Wannier model \ \cite{Falko,
Falko2}. The ground-state solution for a trion was obtained with the Jastrow
form trial function which was optimized using the variational Monte Carlo
method. The theoretical results are in good agreement with experimental
results for the trion.

Binding energies of X$^{+}$ and X$^{-}$ trions were also obtained by mapping
the three-body problem in 2D onto a one particle problem in a
higher-dimensional space \cite{Ganchev}. The interaction (\ref{RKeldysh})
between charge carriers was approximated by the logarithmic potential. The
wave function of three logarithmically interacting particles with different
masses was found by solving the corresponding Schr\"{o}dinger equation. The
resulting binding energies, calculated for various electron-hole mass ratios
were compared with results of the trion binding energies calculated using
the diffusion quantum Monte Carlo method. The comparison of the results
shows that these two theoretical approaches give very close values for the
trion binding energies.

The path integral Monte Carlo method has been used to study the dependence
of trion binding energies on the dielectric screening strength \cite{Saxena}%
. In particular, trion binding energies were investigated for a range of 2D
screening lengths, including the limiting cases of very strong and very weak
screening. One can use these results in the analysis of experimental data
and benchmarking of theoretical and computational models.

The stochastic variational method is applied to excitonic formations of two-
to six-particle systems within semiconducting TMDC using a correlated
Gaussian basis \cite{VargaPRB2016}. The authors studied effects of
electron-hole effective mass ratio as well as the material-specific
effective screening length on the binding energies of these excitonic
formations. Exciton complexes are studied by means of combining the density
functional theory with the path integral Monte Carlo method in order to
accurately account for the particle-particle correlations and the effect of
dielectric environment on the binding energies of excitons, trions and
biexcitons \cite{DenFuncTheoryPIMC}. It was found that the binding energy of
the trion depends significantly stronger on the dielectric environment than
that of biexciton. A comparison of the results \cite{VargaPRB2016,
DenFuncTheoryPIMC} with similar theoretical effective mass model studies
\cite{Saxena, BerkelbachDifMonteCarlo} as well as existing experimental
binding energies for the cases of the exciton and trion show good agreement.
The collinear structure of the trion was used \cite{Thilagam1997} to derive
a simple relation related to the ratio of the trion to exciton binding
energies as which compares well with using the variational quantum Monte
Carlo approach \cite{Ronnow}.

The binding energies of excitons and trions in TMDC monolayers are
investigated using both a multi-band model, taking into account the full
low-energy dispersion for monolayer TMDCs including spin-orbit coupling, and
a single-band model \cite{ZareniaPeeters2017}. Starting with the effective
low-energy single-electron Hamiltonian (\ref{TMDC}) the exciton and trion
Hamiltonians are constructed. To determine the eigenvalues and
eigenfunctions the resulting differential equation is self-consistently
solved using the finite element method. For the single-band model the Schr%
\"{o}dinger equation (\ref{Relative3}) is solved using both the finite
element method and the stochastic variational method in which a variational
wave function is expanded in a basis of a large number of correlated
Gaussians. Reasonable agreement is obtained between the results of both
methods as well as with theoretical studies in the single-band model using
ground-state diffusion Monte Carlo \cite{BerkelbachDifMonteCarlo, Falko} and
path-integral Monte Carlo \cite{DenFuncTheoryPIMC}. However, for the trion
the single-band finite element method results show poor agreement with the
single-band stochastic variational method calculations.

To obtain a solution of the Schr\"{o}dinger equation (\ref{Relative3}) for
the trion in Refs. \cite{KezFBS2017, KezFil2018} the HH method is used by
employing hyperspherical coordinates in 4D configuration space. \ One can
introduce in 4D space the hyperradius $\rho =\sqrt{x_{i}^{2}+y_{i}^{2}}$ and
a set of three angles $\Omega _{i}\equiv (\alpha _{i},\varphi
_{x_{i}},\varphi _{y_{i}}),$ where $\varphi _{x_{i}}$ and $\varphi _{y_{i}}$
are the polar angles for the Jacobi vectors $\mathbf{x}_{i}$ and $\mathbf{y}%
_{i},$ respectively and $\alpha _{i}$ is an angle defined as $x_{i}=\rho
\cos \alpha _{i},$ $y_{i}=\rho \sin \alpha _{i}.$ Using these coordinates
Eq. (\ref{Relative3}) can be rewritten as \cite{KezFBS2017}

\begin{equation}
\left[ -\frac{\hbar ^{2}}{2\mu }\left( \frac{\partial ^{2}}{\partial
^{2}\rho }+\frac{3}{\rho }\frac{\partial }{\partial \rho }-\frac{\widehat{K}%
^{2}(\Omega _{i})}{\rho ^{2}}\right) +\sum_{i>j}^{3}V_{ij}(\left\vert
\mathbf{r}_{i}-\mathbf{r}_{j}\right\vert )-E\right] \Psi (\rho ,\Omega
_{i})=0,\ \   \label{HHtrion}
\end{equation}%
where $\widehat{K}^{2}(\Omega _{i})$ is the angular part of the Laplace
operator in 4D configuration space known as the grand angular momentum
operator \cite{Avery, Jibuti2}.

One can expand the wave function of the trion $\Psi (\rho ,\Omega _{i})$ in
terms of the antisymmetrized HH $\Phi _{K\lambda }(\Omega _{i})$, which are
constructed using the eigenfunctions of the operator $\widehat{K}^{2}$ and
present a complete set of orthonormal basis

\begin{equation}
\Psi (\rho ,\Omega _{i})=\rho ^{-3/2}\sum_{_{K\lambda }}u_{K\lambda }(\rho
)\Phi _{K\lambda }(\Omega _{i}).\   \label{ExpanTrion}
\end{equation}%
In Eq. (\ref{ExpanTrion}) $u_{K\lambda }(\rho )$ are the hyperradial
functions\ and for ease of notation we use $\lambda \equiv $ $\{s,\tau
,l_{x},l_{y},L,M\},$ where $s$ and $\tau $ are a spin and a valley index of
the particle, respectively, $L$ is the total orbital angular momentum of the
trion with $M$ as its projection, $K=2n+l_{x}+l_{y}$, $n$ $\geqslant 0$ is
an integer. By substituting (\ref{ExpanTrion}) into (\ref{HHtrion}) one gets
a set of coupled differential equations for the hyperradial functions $%
u_{K\lambda }(\rho )$ \cite{KezFBS2017}:

\begin{equation}
\left[ \frac{d^{2}}{d\rho ^{2}}-\frac{(K+1)^{2}-1/4}{\rho ^{2}}+\kappa ^{2}%
\right] u_{K\lambda }(\rho )=\frac{2\mu }{\hbar ^{2}}\sum_{_{K^{^{\prime
}}\lambda ^{^{\prime }}}}\mathcal{W}_{K\lambda K^{^{\prime }}\lambda
^{^{\prime }}}(\rho )u_{K^{^{\prime }}\lambda ^{^{\prime }}}(\rho ),
\label{TrionGeneral}
\end{equation}%
where $\kappa ^{2}=2\mu B_{T}/\hbar ^{2}$, $B_{T}$ is the binding energy of
a 2D trion and the coupling effective potential energy is

\begin{equation}
\mathcal{W}_{K\lambda K\lambda ^{^{\prime }}}(\rho )=\int \Phi _{K\lambda
}^{\ast }(\Omega _{i})\sum_{i<j}^{3}V_{ij}(\left\vert \mathbf{r}_{i}-\mathbf{%
\ r}_{j}\right\vert )\Phi _{K^{^{\prime }}\lambda ^{^{\prime }}}(\Omega
_{i})d\Omega _{i},  \label{W3general}
\end{equation}%
which is defined by averaging of the RK potential using the hyperspherical
harmonics $\Phi _{K^{^{\prime }}\lambda ^{^{\prime }}}(\Omega _{i})$ which
fully antisymmetrized with respect to two electrons or two holes and valley
index $\tau =\pm 1$ for X$^{-}$ and X$^{+}$, correspondingly. By solving the
system of hyperradial equations (\ref{TrionGeneral}) numerically one obtains
the corresponding wave function and binding energy for the trions. The
effective potential (\ref{W3general}) is written in the most general form
and one concludes that both the RK and Coulomb potentials can be used for
calculations of binding energies of trions in 2D monolayers.

Results of calculations of the trion binding energies with the RK potential
within HH method \cite{KezFBS2017, KezFil2018} are in good agreement with
similar theoretical effective mass model findings for the trions in MoS$_{2}$%
, MoSe$_{2}$, WS$_{2}$, and WSe$_{2}$ monolayers using the time-dependent
density-matrix functional theory \cite{TimeDepdensity matrix functional
theory}, the stochastic variational method using the explicitly correlated
Gaussian basis \cite{VargaPRB2016}, the path integral Monte Carlo method
\cite{Saxena}, the approach combining density functional theory with the
path Monte Carlo method \cite{DenFuncTheoryPIMC} and the diffusion Monte
Carlo approach  \cite{BerkelbachDifMonteCarlo, Falko}, where the RK
potential was used. Moreover, for trions in TMDC materials there is good
agreement between theory and experiment. Binding energies of trions (X$^{-}$
and X$^{+}$) are sensitive to the effective masses of electrons and holes as
well as to values of the screening length. The reported values of screening
length and effective masses used in some calculations of binding energies
are not necessarily obtained using the same method. The calculations in Ref.
\cite{KezFil2018} demonstrate that for the same screening length the binding
energies of X$^{+}$ slightly exceed the binding energies of X$^{-}$, when
the effective mass of the electron is less than the mass of hole. However,
results also show stronger dependence of binding energies on the screening
length $\rho _{0}$. The latter in some cases can lead to similar binding
energies of X$^{-}$ and X$^{+}.$ The same tendency is also reported in Ref.
\cite{Falko2}, whereas others have found that the binding energy of X$^{-}$
in WSe$_{2}$ monolayer is greater by 10 meV than that for the X$^{+}$ \cite%
{CourtadeSemina}. Statistically exact diffusion quantum Monte Carlo
binding-energy data for a Mott-Wannier model of trions, and biexcitons in 2D
TMDC monolayers in which charge carriers interact via the RK potential
reported in Ref. \cite{Falko2}, confirmed all previous calculations with the
Rytova-Keldysh interaction.

In the framework of the effective mass approximation trions formed by
electrons and holes in layered 2D-dimensional semiconductor heterostructures
were studied as well. It was found that trions in 2D heterostructures are
distinguished by the location of the electrons and holes: the trion is
either formed by an exciton in one layer and an electron (X$^{-}$) or hole (X%
$^{+}$), which is confined in the other layer (direct exciton interacts with
electron or hole from another layer) \cite{KezIJMPB2014} or by two
like-charge particles confined to the same layer and the third (opposite
sign) charge particle confined to another layer (indirect exciton interacts
with electron or hole) \cite{Bondarev2017}, and in both cases the resultant
trions are Coulomb bound. Recent experimental measurements \cite%
{Kimgroup2018} are in good agreement with the calculated binding energy of
the trions formed by indirect excitons in bilayer TMDC structures \cite%
{Bondarev2017}. Variational and diffusion quantum Monte Carlo calculations
of the binding energies and stability of trions in coupled quantum wells
modeled by 2D bilayers with the Coulomb interaction between charge carriers
enables determination of the critical layer separation at which trions
become unbound for various electron-hole mass ratios \cite{Drummond2018}.

\subsection{1D trions}

Quantum wires and carbon nanotubes offer a medium where electrons and holes
are free to move in only one spatial dimension and allow the excitation of
trions. A simple model variational function with a few variable parameters
is proposed for an adequate unified description of X$^{+}$ and X$^{-}$ over
the entire range of free parameters, including the electron-hole mass ratios
and the size and shape of nanowires \cite{QW1 Baars, Zimmermann, Tsu2001,
1DEXP, ExQW, PeetersPRB05, PeetersFewBody2006, PeetersPRB2008, Semina,
Bartnik, Piermarocchi, QW 2012 Phys Lett} and carbon nanotubes \cite%
{Pedersen, Pedersen2, Kammerlander, Pedersen3, DifMC,Matsunaga, Santos,
Bondarev, Watanabe, Pedersen4, ColombierPRL, Yuma, Bondarev2, Bondarev3}. We
cited these articles, but the recent literature on the subject is not
limited to them. In most theoretical approaches the full Hamiltonian of
trions in a NWR is constructed within $k\cdot p$ theory using the
single-band effective mass approximation. The procedure of calculating
effective interaction potentials between charge carriers is identical to the
one discussed in subsection 2C. The one-dimensional Schr\"{o}dinger equation
for 1D trions usually solved within the variational approach \cite{Semina}
or the problem is reduced to the numerical solution of a 2D differential
equation using the finite difference method \cite{PeetersPRB2008, Bartnik,
QW 2012 Phys Lett}. Theoretical calculations have been carried out to
investigate the binding energies of the trions in NWRs \cite%
{PeetersFewBody2006, PeetersPRB2008}. The binding energy for charged
excitons X$^{-}$ and X$^{+}$is calculated within the single-band effective
mass approximation including effects due to strain for different confinement
geometry of NWRs and dependence of the trion binding energy on the size and
shape of the NWR was investigated numerically. Surprisingly, in the
Born-Oppenheimer approximation the Schr\"{o}dinger equation for 1D X$^{+}$
trion and biexciton in core/shell NWR can be solved analytically for a
cusp-type Coulomb interaction and one can obtain analytical expressions for
the binding energies and wavefunctions \cite{RK1D}. Theoretical studies
confirm that both the lateral confinement and the localization potential
have a strong effect on the relative stability of the trions in NWRs. Even a
weak localization potential not only enhances the binding energy but also
changes the relative stability of the positive and negative trions \cite{QW
2012 Phys Lett}. Trions in cylindrical nanowires with a dielectric mismatch
within the adiabatic approximation is investigated and the three-particle
problem reduces to an effective two-dimensional Schr\"{o}dinger equation
\cite{Slachmuylders1} and one-dimensional Schr\"{o}dinger equation \cite%
{Slachmuylders2} for the relative motion, which are solved numerically.
Analysis of the results for the singlet and triplet trion binding energies
shows that the X$^{-}$ is always less stable than the X$^{+}$ in a wire with
hole to electron mass ratio more than 1 \cite{Slachmuylders1}. It is
demonstrated that the dielectric mismatch effects result in a distorted
Coulomb interaction between the charge carriers \cite{Slachmuylders2}.
Recently, a first-principles approach to trion excitations based on an
extension of the Bethe-Salpeter equation to three-particle states was
developed and applied to carbon nanotubes \cite{Deilmann2016}. The method
provides detailed insights into the physical nature of trion states and
application for a semiconducting (8,0) CNT shows that the optically active
trions are redshifted by $\thicksim $130\thinspace \thinspace meV compared
to the excitons, which confirms experimental findings for similar CNTs. A
configuration space method for calculating binding energies of exciton
complexes in carbon nanotubes, as well as a review on this subject, is given
in Ref. \cite{Bondarev3}.

\section{Four Body Problem$-$Biexcitons}

Experimental and theoretical studies of bound biexcitons in bulk and
low-dimensional semiconductors have greatly advanced our fundamental
understanding of few-body physics in semiconductors. A biexciton is an
excitonic complex consisting of two electrons and two holes, where the
charge carriers are bound via electromagnetic interaction, and have been
observed in 3D, 2D, and 1D semiconductor structures. Due to computational
difficulties for the description of this system, more work has been done on
trions than on biexcitons. Most of theoretical approaches to study
biexcitons with pure Coulomb interaction are variational and using the
sophisticated technique borrowed from atomic and molecular physics.\ Some
approaches attempt to reduce the four-particle Hamiltonian to that of a
smaller system with fewer degrees of freedom \cite{Lozovik1D}. Both the pure
Coulomb and screened Coulomb potentials are used to describe the interaction
of the biexciton in low-dimensional semiconductors. The most common
approaches for solving the biexcitonic system include the variational
method, diffusion Monte Carlo, and HH.

\subsection{3D biexcitons}

The first reported observation of biexcitons in GaAs QWs came in 1982 \cite%
{MillerQW1982}, eleven years before the first evidence of trions in quantum
wells \cite{QW1}. A large number of papers concerning different aspects of
quantum well biexcitons has been published. The first variational
calculations of the ground state of 3D and 2D biexcitons was performed by
considering a biexciton as two weakly interacting subsystems. The special
coordinate transformation reduces the four-body problem to the problem of
one quasiparticle weakly interacting with a rigid core of two strongly
interacting quasiparticles\textit{. }This model yielded too large binding
energies of 3.26 eV and 0.354 eV for a 2D and 3D biexciton, respectively.
Along with variational approaches with different trial wave functions, the
stochastic variational method with correlated Gaussian basis \cite{Varga1999}
was used to study the binding energies and other properties of the
biexcitons. To solve the Schr\"{o}dinger equation for four charged particles
interacting via the Coulomb potential, the trial wave functions are chosen
to be combinations of correlated Gaussians functions and the stochastic
variational method has been used as the most adequate choice of the
nonlinear parameters of the correlated Gaussians. The correlated Gaussians
allow a fully analytical calculation of the matrix elements. The stability
of a system of two positively and two negatively charged particles with
unequal masses in 3D was studied by means of a variational Monte Carlo
optimization and quantum diffusion Monte Carlo methods \cite{Bressanini1998}%
. In the case of 3D biexcitons it was shown that biexciton is stable against
the dissociation in two excitons for heavy hole masses. These methods have
also provided upper and lower bounds to the binding energies of 2D
biexcitons.

\subsection{2D biexcitons}

Biexcitons have been studied in QWs since their discovery in 1982. However,
here let us focus on formation of biexcitons in monolayers of 2D materials.
In TMDC monolayers stable bound states of biexcitons were reported in Refs.
\cite{ShangBiexiton, WS2Plechinger, MaiBiexiton, Reichman2015,
SieBiexitonMoS2, MoSe2Hao}. The observed binding energies in monolayers of
MoS$_{2}$, WS$_{2}$ and WSe$_{2}$ are in range of $\backsim $ $40-70$ meV.
In particular, the measured binding energies of biexcitons in TMDC are the
following: 40, 60 meV \cite{SieBiexitonMoS2}, 70 meV \cite{MaiBiexiton} (MoS$%
_{2})$; 45 meV \cite{ShangBiexiton}, 65 meV \cite{WS2Plechinger} (WS$_{2})$;
52 meV \cite{Reichman2015} (WSe$_{2})$. A recent study \cite{MoSe2Hao} which
used resonant two-dimensional coherent spectroscopy has identified a
biexciton in MoSe$_{2}$ with binding energy of $\thicksim 20$ meV, and also
observed the charged bound biexciton with a binding energy of 5 meV.
Therefore, the binding energy of the biexciton in MoSe$_{2}$ is
significantly smaller compared to previously reported experimental values
for other TMDC monolayers. One possible explanation is that previous
experiments have observed charged biexcitons or excited-state biexcitons,
because it was not possible to distinguish different types of higher-order
bound states based on the one-dimensional spectroscopy methods used in the
previous studies \cite{MoSe2Hao}.

The same theoretical approaches, which were developed for investigation of
trions in TMDC materials, are used in studies of biexcitons. In particular,
2D biexcitons have been studied in the framework of the Mott-Wannier model
using quantum Monte Carlo methods, variational methods, and the method of
hyperspherical harmonics. To find biexciton binding energies in TMDC
monolayers using the variational method, intricate trial wave functions were
employed \cite{Reichman2015, Prada}. A variational calculation \cite%
{Reichman2015} of the biexciton state reveals that the high binding energy
arises not only from strong carrier confinement, but also from reduced and
non-local dielectric screening. Biexcitons in low dimensional TMDCs were
also studied in the framework of the stochastic variational method using the
explicitly correlated Gaussian basis \cite{VargaNano2015, VargaPRB2016}.
Within the effective mass approach, quantum Monte Carlo methods, such as the
diffusion Monte Carlo and the path integral Monte Carlo, provide accurate
and powerful means for studying few-particle systems. Biexcitons in 2D TMDC
sheets of MoS$_{2}$, MoSe$_{2}$, WS$_{2}$, and WSe$_{2}$ are studied by
means of the density functional theory and path integral Monte Carlo method
in \cite{DenFuncTheoryPIMC}. The diffusion Monte Carlo method provides a
useful approach for studying the energetics of excitonic complexes. This
approach was used to investigate the binding energies and intercarrier
radial probability distributions of biexcitons in a variety of TMDC
monolayers \cite{BerkelbachDifMonteCarlo}. The binding energies of
biexcitons in TMDC monolayers studied within the framework of a
nonrelativistic potential model using the method of hyperspherical harmonics
in six-dimensional configuration space for solution of a four-body Schr\"{o}%
dinger equation \cite{KezFBS2017}. In all of these studies, the
Rytova-Keldysh potential was used to account for the strong, non-local
screening and it was found that the binding energies of biexcitons in
TMDCs monolayers are less than trions.

The comparison of the results of \ calculations of the binding energies of
biexciton in TMDC monolayers obtained by using the Rytova-Keldysh potential
\cite{Keldysh} shows good agreement between a variety of approaches,
including studies in the framework of the stochastic variational method
using a correlated Gaussian basis \cite{VargaNano2015, VargaPRB2016},
theoretical studies in the single-band model using ground-state diffusion
Monte Carlo \cite{BerkelbachDifMonteCarlo, Falko} and density functional
theory and path integral Monte Carlo \cite{DenFuncTheoryPIMC} methods and
method of hyperspherical harmonics \cite{KezFBS2017}. In average, the
discrepancies are less than $\pm $1 meV. There is a discrepancy with
experiment for the biexciton case for MoS$_{2}$, WS$_{2}$, and WSe$_{2}$
with all theoretical predictions, while the recent experimental result for
MoSe$_{2}$ \cite{MoSe2Hao} is in reasonable agreement with theoretical
calculations. Binding energies of excitonic complexes examined using the
fractional dimensional space approach \cite{Thilagam1997}. The binding
energies of the exciton, trion, and biexciton in TMDCs of varying layers are
analyzed, and linked to the noninteger dimensionality parameter $\alpha $.
Estimates of the binding energies of exciton complexes for the monolayer
configuration of TMDC suggest a non-collinear structure for the trion and a
positronium-molecule-like square structure for the biexciton \cite%
{Thilagam2014}.

In Ref. \cite{Falko} the diffusion quantum Monte Carlo approach is applied
to find the energies for Mott-Wannier models of trions and biexcitons in
monolayer 2D TMDC semiconductors. Calculations are performed by using the RK
potential for the interaction between charge carriers. Calculations indicate
that the binding energy of a trion is larger than the biexciton binding
energy in 2D semiconductors. Moreover, trion binding energies are
significantly more sensitive to the effective mass values of electrons and
holes than biexciton binding energies. Results of this study suggest that
the experimental \textquotedblleft trion\textquotedblright\ and
\textquotedblleft biexciton\textquotedblright\ peaks may be misclassified,
because the trion binding energy should exceed the biexciton binding energy,
but also indicate that the RK potential fails to give a quantitative
description of the observed excitonic properties of 2D TMDCs \cite{Falko}.

In summary, a broad range of theoretical works \cite{VargaNano2015,
VargaPRB2016, BerkelbachDifMonteCarlo, Falko, DenFuncTheoryPIMC, KezFBS2017,
Falko2} on 2D biexciton binding energies show excellent quantitative
agreement with each other, but an enormous, two to threefold disagreement
with experiment. Only for MoSe$_{2}$ there is reasonable agreement with the
experimental measurement of binding energy reported in Ref. \cite{MoSe2Hao}.

Within considered approaches of the treatment of biexcitons the origin of a
discrepancy between experimental observations and theoretical calculations
from theory point of view could only arise for one or more of the following
reasons: i. The Mott--Wannier model is incorrect or incomplete; ii. The
Mott--Wannier model is in principle correct, but the screening considered
via the RK potential requires the modification of this potential; iii. The
effective masses of electrons and holes or other parameters used as inputs
in the models are incorrect. There is no obvious reason to believe that the
Mott--Wannier model, which provides a good description of excitons and a
reasonable explanation for the binding energy of trions is incorrect. May be
the discrepancy indicates that the description with the RK potential is
still lacking some features for the TMDC class of materials? Perhaps a
recent derived new potential \cite{TuanNewRKPot} for the charge carrier
interaction in a TMDC monolayer, which takes into account the three atomic
sheets that compose a TMDC monolayer, can address this problem. However, the
most likely explanation for the disagreement with experiments is a
misinterpretation or misclassification of experimental optical spectra \cite%
{Falko, Falko2}. Indeed, last year a set of four papers by different groups
were published together in Nature Communication \cite{Chen2018,
Waldecker2018, Li2018, Barbone2018}, that clarify this disagreement: the
previously observed "biexcitons" in WSe$_{2}$ \cite{Reichman2015} turned out
to be a charged biexciton state. These observations were made possible by
large advancements in sample growth and fabrication of a high-quality
single-layer WSe$_{2}$. In Ref. \cite{Chen2018} \ a biexciton in WSe$_{2}$
was observed with a binding energy of 18 meV. The authors of \cite%
{Waldecker2018} are reported the binding energy about 20 meV, while the
experimental study based on low-temperature photoluminescence spectroscopy
\cite{Li2018} reported the binding energy of the biexcitons in the
BN-encapsulated single-layer WSe$_{2}$ to be about 16--17 meV. These
experimental measurement are in reasonable agreement with the theoretical
predictions \cite{VargaNano2015, VargaPRB2016, BerkelbachDifMonteCarlo,
Falko, DenFuncTheoryPIMC, KezFBS2017, Falko2}. Thus, the discrepancy of the
biexciton binding energy found between previous experiments and theories is
resolved: the experimental and calculated biexciton binding energy is indeed
smaller than the trion binding energy due to the screened charge carrier
potential in 2D TMDCs. To investigate the biexciton valley configuration is
was applied a strong out-of-plane magnetic field, which acts to break the
valley degeneracy. Interestingly enough, it was reported that the biexciton
in WSe$_{2}$ consists of a spin-zero bright exciton in one valley and a
spin-one \textquotedblleft dark\textquotedblright\ exciton in the other
which is unusual configuration of the exciton molecule. In contrast to Refs.
\cite{MoSe2Hao} and \cite{Steinhoff2018}, where 20 meV biexcitons in MoSe$%
_{2}$ and WSe$_{2}$ were observed using two-dimensional coherent
spectroscopy and ultrafast pump-probe spectroscopy, respectfully, and both
works describe bright--bright excitons, which would suggest the biexciton
binding energy is only weakly sensitive to the spin configuration of the two
constituent excitons \cite{Waldecker2018}.

In Ref. \cite{Barbone2018} was determined the bound state of the electrons
and holes comprising the biexcitons through magneto-optical spectroscopy and
also resolved a splitting of 2.5 meV for the biexciton in WSe$_{2}$, which
was attribute to the fine structure. In Ref. \cite{Steinhoff2018} is also
shown that biexcitons in monolayer TMDCs exhibit a distinct fine structure
on the order of meV due to electron-hole exchange. Experiments on monolayer
WSe$_{2}$ reveal decisive biexciton signatures and a fine structure in
excellent agreement with a microscopic theory that shows that the biexciton
fine structure is caused by nonlocal electron-hole exchange, while local
exchange leads to an increase of the binding energy of the lowest biexciton
state.

Biexcitons that are formed by the indirect excitons in heterostructures
modeled by 2D bilayers are studied within the effective mass approximation
with the Coulomb interaction between charge carriers. A number of
theoretical approaches can be found in the literature \cite{Pack1970,
Tan2005, Fogler2008, Bondarev2017, Drummond2018}. The binding energy and
wave functions of two-dimensional indirect biexcitons are studied
analytically and numerically using the stochastic variational method \cite%
{Fogler2008}. It is proven that stable biexcitons exist only when the
distance between electron and hole layers is smaller than a certain critical
threshold. Variational and diffusion quantum Monte Carlo calculations of the
binding energies of isolated indirect trions and biexcitons in ideal
two-dimensional bilayer systems within the effective mass approximation were
performed in Ref. \cite{Drummond2018}. The authors have found that for
indirect trions, the critical layer separation at which the trion becomes
unbound is at least an order of magnitude larger than that of the biexciton
and concluded that in 2D materials the binding energy of the trion relative
to the biexciton is further magnified by the nonlocal screening of the
charge carriers by the 2D layers.

Let us also mention that studies of the exciton complexes can be extended
beyond of the biexciton. Theoretical studies have predicted the existence of
numerous multi-particle excitonic states. However, more complex states
beyond of biexcitons have been elusive due to limited spectral quality of
the optical emission. As it is mentioned above authors of Ref. \cite{MoSe2Hao} have identified the charged bound biexciton in MoSe$_{2}$ with a binding energy of 5 meV. The recent experimental studies \cite{Chen2018,
Waldecker2018, Li2018, Barbone2018} reported five-particle valleytronic
states in an atomically-thin WSe$_{2}$ semiconductor $-$ quintons $-$
negatively or positively charged biexcitons, when one free electron or hole
binds to a biexciton. In particular, these studies reported the
identification of negatively charged biexcitons formed from a trion X$^{-}$
and a neutral exciton. The observed negatively charged quinton XX$^{-}$
binding energy depends on spin-valley configurations, and is in agreement
with theoretical calculations \cite{DenFuncTheoryPIMC}. In contrast to the
negatively charged trions, no fine features can be resolved in the XX$^{-}$
peak at this stage \cite{Waldecker2018}. The ground and excited states of
exciton-trions are predicted to be bound and their structures are
investigated within different theoretical approaches \cite%
{DenFuncTheoryPIMC, VargaPRB2016, Steinhoff2018}. What about the excitonic
complexes beyond quintons? One such study \cite{VargaPRB2016} considered up
to six particles in TMDC\ monolayers. The authors applied the stochastic
variational method using an accurate correlated Gaussian basis to calculate
the energies for two- to six-body excitonic formations by employing the RK
potential.

The study of biexcitons is not limited by TMDC monolayers, but extends to
the emerging anisotropic 2D semiconductors such as phosphorene, which shows
strongly anisotropic optical and electrical properties. This anisotropy
leads to the formation of quasi-1D biexciton in a 2D system, which results
in even stronger many body interactions in anisotropic 2D materials, arising
from the further reduced dimensionality of the quasi-particles and thus
reduced dielectric screening. Within an effective mass theory using
diffusion Monte Carlo method, biexcitons in anisotropic 2D materials are
investigated in Ref. \cite{ReichmanPhosforos}, where the presented binding
energies biexcitons in phosphorene and arsenene are notably larger than
those for TMDC monolayers. In particular, for monolayer phosphorene the
binding energy is twice as large as it for a typical TMDC monolayer.

\subsection{1D biexcitons}

While biexcitons were originally identified in QWs in 1982 \cite%
{MillerQW1982}, it would take twenty years before the first direct
observation of the 1D biexciton was reported in a high quality semiconductor
quantum wire with a binding energy of 1.2 meV \cite{Crottini1D}. Biexcitons
in NWRs and CNTs are studied using similar methods \ as for the trions. A
study of a series of NWRs with aspect ratios (length over diameter) ranging
from 1 to 10 shows that the multiexciton generation rates are roughly
independent of the NWR diameter \cite{Baer2013}. Taking into account the
behavior of the biexciton binding energy with the NWR size variation, it was
proposed that there exists an optimal radius of elongated quantum wire, for
which the associative ionization of biexciton antibonding states into trion
bonding states occurs that leads to the formation of trions \cite{RK1D}.
Carbon nanotubes, due to their strongly diameter-dependent excitonic binding
energy, can exhibit quasi-1D properties. Moreover, electron and hole
effective masses depend on the diameter and chirality of the CNT, which also
affects the binding energies of excitonic complexes. Different methods have
been applied to study the 1D biexciton in CNTs. Based on the method widely
used in atomic physics \cite{Herring1964} an analytical expression for the
binding energy of the biexciton in a small-diameter CNT is obtained as a
function of the interexciton distance and binding energy of constituent
quasi-1D excitons in carbon nanotubes \cite{Bondarev}. The latter allows one
to trace biexciton energy variation, whereby the exciton binding energy
varies. Applying the tight-binding model to calculate the binding of
biexcitons the corresponding Coulombian 1D Schr\"{o}dinger equation for four
charged particles was solved by using the quantum Monte Carlo \cite%
{Kammerlander} and variational \cite{Pedersen2} approaches. For typical
nanotube diameters, biexciton binding energies obtained using the quantum
Monte Carlo method are much larger than predicted by the variational method.
Due to their large binding energies, biexcitons in CNTs might be stable
against thermal fluctuations at room temperature.

It is of particular interest to apply a few-body methods, such as the
hyperspherical harmonics expansion in one dimension formulated in terms of
an expansion on a single-particle oscillator basis suggested in \cite%
{Timofeyuk}, for description of 1D trions and biexcitons.

\section{Conclusions}

This review has discussed some of the concepts, theoretical approaches, and
computational methods used to describe excitons, trions and biexcitons in
three-, two- and one-dimensional configuration spaces in various types of
materials. It is shown that the reduction of dimensionality generally
enhances the binding energies of exciton complexes, leading to a host of
possible novel applications for experimentalists to explore. At the same
time, changing the state space configuration from 3D to 2D to 1D inevitably
introduces new theoretical challenges, some of which have been solved by
clever application of highly specialized mathematical techniques, while for
the time being, some challenges have only been tackled by leveraging
computational techniques.

Quantum confinement and the lack of bulk dielectric screening have profound
effects on the binding energies of excitons, trions and biexcitons in
low-dimensional semiconductors. While in bulk materials, the interaction
between charge carriers is weak due to the dielectric screening given by a
simple multiplicative renormalization by the dielectric constant, the highly
non-local nature of the dielectric screening in 2D and 1D materials such as
atomically thin crystals and nanowires is responsible for the dramatic
increase the binding energies, as well as for a number of unique properties
exhibited by the 2D and 1D materials. Calculations demonstrated that the
large binding energy of excitons, trions and biexcitons in 2D and 1D
materials is mostly a result of weak dielectric screening rather than
quantum confinement. It is illustrated that the binding energies of
excitonic complexes in 2D and 1D materials are usually distinctly different
from those of their 3D counterparts. One should mention that reducing the
dimensionality of a system is often associated with exceptional electronic,
optical, and magnetic properties, as the reduction of available phase space
and diminished screening lead to enhanced quantum effects and increased
correlations \cite{Kim2016}.

There are a number of issues which should be addressed. Today we have a
number of theoretical studies of 2D trions using a variety of theoretical
approaches. It would be great to have a comprehensive study of trions fine
structure in 2D monolayers within the method of the Faddeev equation. The
latter requires an extension of this formalism to two-dimensional
configuration or momentum spaces with proper consideration of non-local
screening of the Coulomb interaction and the coupled spin and valley
pseudospin degrees of freedom.

Biexcitons were studied using different methods such as a variational,
diffusion Monte Carlo and HH. However, there is no comprehensive study of
biexcitons within the Faddeev-Yakubovsky formalism in 3D. Moreover, there is
a lack of the Faddeev-Yakubovsky formalism either in configuration or
momentum 2D spaces. Moreover, a complete understanding of multi-exciton
complexes is key to study coherent many-body phenomena, such as condensation
\cite{Kogar2017}, superconductivity \cite{Cotlet2016} and superfluidity \cite%
{Fogler}, requires a solution of five- and six-body problem for electron-hole
system in 2D materials.

It is of particular interest to consider trions and biexcitons in 2D
materials such as phosphorene, which demonstrates a strong anisotropic
nature, in the framework of the Faddeev and Faddeev-Yakubovsky equations and
method of HH. This requires an extension of these methods for the
description of three and four particles with anisotropic masses in $x$- and $%
y$-direction. The study the binding energies of direct and indirect trions
and biexcitons in monolayers and double layer heterostructures of Xenes in an
external electric field applied perpendicular to the plane of the Xene
monolayer is also an important task. The external electric field changes the
band gap of the Xenes monolayers and thus the effective masses of electrons
and holes. This allows to study how the external electric field can be used
to tune the bicentenaries and other properties of trions and biexcitons by
changing the effective mass of charge carriers.

A special concern is related to a double layered van der Waals
heterostructures of 2D materials. The computational modeling of two
monolayer heterostructures separated by a dielectric is complicated by the
incommensurable nature of the interfaces. Consequently, reliable modeling
of realistic, incommensurable heterostructures requires development of novel
approaches that combine the quantum description of the individual layers
with a more coarse grained description of the effect of interlayer
interactions \cite{ThygesenReview2017}.

A comprehensive study of formation of trions and biexcitons in bilayer
heterostructures of 2D materials in addition to contemporary calculations
needs development and application of the modelless approaches for restricted
3D space (the motion of electrons and holes is frozen in $z$-direction)
based on the methods of Faddeev and Faddeev-Yakubovsky equations, method of
HH and the variational method with the explicitly correlated Gaussian basis
functions in restricted 3D space.

\section*{Acknowledgements}

This work is supported by the U.S. Department of Defense under Grant No.
W911NF1810433 and PSC CUNY under Grant No. 62261-00 50.


\begin{thebibliography}{999}
\bibitem{TerMartiros} \label{2body}G. V. Skorniakov and K. A.
Ter-Martirosian, ZhETF \textbf{31}, 775 (1957) [Sov. Phys. JETP \textbf{4},
648 (1957)].

\bibitem{Faddeev} L. D. Faddeev, ZhETF \textbf{39}, 1459 (1961) [Sov. Phys.
JETP \textbf{12}, 1014 (1961)].

\bibitem{Eyges} L. Eyges, Phys. Rev. \textbf{115}, 1643 (1959).

\bibitem{Gribov} V. N. Gribov, ZhETF \textbf{38}, 553 (1960), [Soviet Phys.
JETP \textbf{11}, 400 (1960)].

\bibitem{FaddeevTrudi} L. D. Faddeev. Mathematical problems of the quantum
theory of scattering for a system of three particles. Proceedings of the
Mathematical Institute of the Academy of Sciences of the USSR. \textbf{69},
1-122 (1963).

\bibitem{Efimov2019} V. Efimov, Few-Body Syst. \textbf{60}, 15 (2019).

\bibitem{Efimov1970} V. Efimov, Yad. Fiz. \textbf{12}, 1080 (1970).

\bibitem{Efimov1970PL} V. Efimov, Phys. Lett. \textbf{33} B, 563 (1970).

%\U{41b}. \U{414}. \U{424}\U{430}\U{434}\U{434}\U{435}\U{435}\U{432}, \U{41c}%
%\U{430}\U{442}\U{435}\U{43c}\U{430}\U{442}\U{438}\U{447}\U{435}\U{441}\U{43a}%
%\U{438}\U{435} \U{432}\U{43e}\U{43f}\U{440}\U{43e}\U{441}\U{44b} \U{43a}%
%\U{432}\U{430}\U{43d}\U{442}\U{43e}\U{432}\U{43e}\U{439} \U{442}\U{435}%
%\U{43e}\U{440}\U{438}\U{438} \U{440}\U{430}\U{441}\U{441}\U{435}\U{44f}%
%\U{43d}\U{438}\U{44f} \U{434}\U{43b}\U{44f} \U{441}\U{438}\U{441}\U{442}%
%\U{435}\U{43c}\U{44b} \U{442}\U{440}\U{451}\U{445} \U{447}\U{430}\U{441}%
%\U{442}\U{438}\U{446}. -- \U{422}\U{440}\U{443}\U{434}\U{44b} \U{41c}\U{430}%
%\U{442}\U{435}\U{43c}\U{430}\U{442}\U{438}\U{447}\U{435}\U{441}\U{43a}\U{43e}%
%\U{433}\U{43e} \U{438}\U{43d}\U{441}\U{442}\U{438}\U{442}\U{443}\U{442}%
%\U{430} \U{410}\U{41d} \U{421}\U{421}\U{421}\U{420}. \textbf{69}. \U{441}%
%.1-122 (1963).

\bibitem{Yakubovsky} O. A. Yakubovsky, Sov. J. Nucl. Phys. \textbf{5}, 937
(1967).

\bibitem{Merkuriev} S. P. Merkuriev, C. Gignoux, and A. Laverne, Ann. Phys.
\textbf{99}, 30 (1976).

\bibitem{MF85} L. D. Faddeev and S. P. Merkuriev, \textit{Quantum Scattering
Theory for Several Particle Systems.} Nauka, Moscow, 1985; Kluwer Academic,
Dordrecht, 398 pp. 1993.

\bibitem{Varga1} Y. Suzuki and K. Varga, Stochastic Variational Approach to
Quantum-Mechanical Few-Body Problems. Springer-Verlag Berlin Heildelberg,
310 pp. 1998.

\bibitem{BubinRMP} J. Mitroy, S. Bubin, W. Horiuchi, Y. Suzuki, L.
Adamowicz, W. Cencek, K. Szalewicz, J. Komasa, D. Blume, and K. Varga, Rev.
Mod. Phys. \textbf{85}, 693 (2013).

\bibitem{Morpurgo} G. Morpurgo, Nuovo Cimento \textbf{9}, 461 (1952).

\bibitem{Delves} L. M. Delves, Nucl. \textbf{9}, 391 (1959).

\bibitem{Delves2} L. M. Delves, Nucl. Phys. \textbf{20}, 276 (1960).

\bibitem{Delves3} L. M. Delves, Nucl. Phys. \textbf{29}, 268 (1962).

\bibitem{Smith} F. T. Smith, Phys. Rev. \textbf{120}, 1058 (1960).

\bibitem{Simonov} Yu. A. Simonov, Yad. Fiz. \textbf{3}, 630 (1966). [Sov. J.
Nucl. Phys. \textbf{3}, 461 (1966)].

\bibitem{Simonov2} A. M. Badalyan and Yu. A. Simonov, Yad. Fiz. \textbf{3},
1032 (1966).

\bibitem{Avery} J. Avery, Hyperspherical Harmonics: Applications in Quantum
Theory (Dordrecht: Kluwer) (1989).

\bibitem{Jibuti2} R. I. Jibuti and K. V. Shitikova, Method of hyperspherical
functions in atomic and nuclear physics Energoatomizdat, Moscow, 270 pp.
1993 (\textit{in Russian}).

\bibitem{Frenkel} J. Frenkel, Phys. Rev. \textbf{37}, 17 (1931).

\bibitem{WannierM} G. Wannier, Phys. Rev. \textbf{52}, 191 (1937).

\bibitem{Novoselov} K. N. Novoselov, A. K. Geim, S. V. Morozov, D. Jiang, Y.
Zhang, S.V. Dubonos, I. V. Grigorieva, and A.A. Firsov, Science \textbf{306}%
, 666 (2004).

\bibitem{Bhimanapati} G. R. Bhimanapati, et al., ACS Nano \textbf{9,} 11509
(2015).

\bibitem{Novoselov2} K. S. Novoselov, A. Mishchenko, A. Carvalho, A. H.
Castro Neto, Science \textbf{353}, 9439 (2016).

\bibitem{Velicky} M. Velick\'{y} and P. S. Toth, App. Mater. Today \textbf{8}%
, 68 (2017).

\bibitem{Jariwala} D. Jariwala, T. J. Marks, \ M. C. Hersam, Nat. Mater.
\textbf{16,} 170 (2017).

\bibitem{Novoselov3} A. K. Geim, K. S. Novoselov, Nat. Mater. \textbf{6},183
(2007).

\bibitem{Kormanyos} A. Korm\'{a}nyos, G. Burkard, M. Gmitra, J. Fabian, V. Z%
%TCIMACRO{\U{b4}}%
%BeginExpansion
\'{}%
%EndExpansion
olyomi, N. D. Drummond, and V. Fal$^{\prime }$ko, 2D Mater. \textbf{2},
022001 (2015).

\bibitem{Joshua} J. O. Island et al., 2D Mater. \textbf{4}, 022003 (2017).

\bibitem{Li1} L. Li, Y. Yu, G. J.Ye, Q. Ge, X. Ou, H. Wu, D. Feng, X. H.
Chen, and Y. Zhang, Nat. Nanotechnol. \textbf{9,} 372 (2014).

\bibitem{Li2} H. Liu, A. T. Neal, Z. Zhu, Z. Luo, X. Xu, D. Tom\'{a}nek, and
P. D.Ye, ACS Nano \textbf{8,} 4033 (2014).

\bibitem{Warren2015} A.~H. Woomer, T.~W. Farnsworth, J. Hu, R.~A. Wells,
C.~L. Donley, and S.~C. Warren, ACS Nano \textbf{9}, 8869 (2015).

\bibitem{Dean2010} C. R. Dean, et al., Nat. Nanotechnol. \textbf{5,} 722
(2010).

\bibitem{Matthes2014} L. Matthes, P. Gori, O. Pulci, and F. Bechstedt, Phys.
Rev. B \textbf{87}, 035438 (2013).

\bibitem{Molle2017} A. Molle, J. Goldberger, M. Houssa, Y. Xu, S. C. Zhang,
and D. Akinwande, Nat. Mater. \textbf{16}, 163 (2017).

\bibitem{Brunetti2018} M. N. Brunetti, O. L. Berman, and R. Ya.
Kezerashvili, Phys. Rev. B \textbf{98}, 125406 (2018).

\bibitem{Falko2012} N. D. Drummond, V. Z\'{o}lyomi, and V. I. Fal$^{\prime }$%
ko, Phys. Rev. B \textbf{85}, 075423 (2012).

\bibitem{Davila} M. E. D\'{a}vila, L. Xian, S. Cahangirov, A. Rubio, and G.
Le Lay, New J. Phys. \textbf{16,} 095002 (2014).

\bibitem{Zhu2014} F-F. Zhu, W-J. Chen, Y. Xu, C-L. Gao, D-D. Guan, C-H. Liu,
D. Qian, S-C. Zhang, and J-F. Jia, Nat. Mater. \textbf{14}, 1020 (2015).

\bibitem{Mannix2015} A. J. Mannix, et al., Science \textbf{350,} 1513 (2015).

\bibitem{Geim2014} A. Geim, I. Grigorieva, Nature \textbf{499}, 419 (2014).

\bibitem{Rytova} N. S. Rytova, Proc. Moscow Stare University, Phys. Astron.
\textbf{3}, 30 (1967).

\bibitem{Keldysh} L.V. Keldysh, JETP Lett. \textbf{29}, 658 (1979). L. V.
Keldysh,\ JETP Lett. \textbf{29}, 658 (1980) [Pis'ma Zh. Eksp. Teor. Fiz.
\textbf{29}, 716 (1979)].

\bibitem{Dery2019} B. Scharf, D. Van Tuan, I. \v{Z}uti\'{c}, and H. Dery, J.
Phys.: Condens. Matt. \textbf{31}, 203001 (2019).

\bibitem{Nishanov} Yu. E. Lozovik and V. N. Nishanov, Sov. Phys. Solid State
\textbf{18}, 1905 (1976).

\bibitem{MoskalenkoSnoke} S. A. Moskalenko and D. W. Snoke, Bose-Einstein
Condensation of Excitons and Biexcitons and Coherent Nonlinear Optics with
Excitons (Cambridge University Press, New York, 2000).

\bibitem{Lozovik} Yu. E. Lozovik and V. I. Yudson, Sov. Phys. JETP. \textbf{%
44}, 389 (1976)\textbf{; }Zh. Eksp. Teor. Fiz. \textbf{71}. 738 (1976).

\bibitem{Snoke} D. Snoke, Spontaneous Bose coherence of excitons and
polaritons, Science \textbf{298}, 1368 (2002).

\bibitem{Butov} L. V. Butov, J. Phys.: Condens. Matter \textbf{16} R1577
(2004).

\bibitem{Combescot} M. Combescot, R. Combescot, and F. Dubin, Rep. Prog.
Phys. \textbf{80}, 066501 (2017).

\bibitem{Hunt2018} M. Danovich, D. A. Ruiz-Tijerina, R. J. Hunt, M.
Szyniszewski, N. D. Drummond, and V. I. Fal$^{\prime }$ko, Phys. Rev.
\textbf{97}, 195452 (2018).

\bibitem{BermanKezerashviliFBS2011} O.~L. Berman and R.~Ya. Kezerashvili,
Few-Body Syst. \textbf{50}, 407 (2011).

\bibitem{RKPhysRevB2012} O. L. Berman, R. Ya. Kezerashvili, and K. Ziegler,
Phys. Rev. B \textbf{85}, 035418 (2012).

\bibitem{RKPhysRevB2016} O.~L. Berman and R.~Ya. Kezerashvili, Phys. Rev. B
\textbf{93}, 245410 (2016).

\bibitem{RKPhysRevB2017} O.~L. Berman and R.~Ya. Kezerashvili, Phys. Rev. B
\textbf{96}, 245410 (2017).

\bibitem{Guinea} A. H. Castro Neto, F. Guinea, N. M. R. Peres, K. S.
Novoselov, and A. K. Geim, Rev. Mod. Phys. \textbf{81}, 109 (2009).

\bibitem{Vozmediano} M. A. H. Vozmediano, M. I. Katsnelson, and F. Guinea,
Phys. Rep. \textbf{496}, 109 (2010).

\bibitem{Xiao} D. Xiao, G. B. Liu, W. Feng, X. Xu, and W. Yao, Phys. Rev.
Lett. \textbf{108}, 094502 (2012).

\bibitem{Tabert} C. J. Tabert and E. J. Nicol, Phys. Rev. B \textbf{89},
195410 (2014).

\bibitem{Landau} L. D. Landau and E. M. Lifshitz, Quantum Mechanics:
Non-Relativistic Theory, 3rd ed. (Elsevier, Oxford, 1977).

\bibitem{Liboff} R. L. Liboff, Introductory Quantum Mechanics, 2nd ed.
(Addison-Wesley, Reading, MA, 1992).

\bibitem{Griffiths} D. J. Griffiths, Introduction to Quantum Mechanics, 2nd
ed. (Prentice-Hall, New York, 2005).

\bibitem{RKPhysRevA2012} O.~L. Berman and R.~Ya. Kezerashvili, and K.
Ziegler, Phys. Rev. A \textbf{87}, 042513 (2013).

\bibitem{Alba} D. Alba, H. W. Crater, and L. Lusanna, J. Phys. A: Math.
Theor. \textbf{40}, 9585 (2007).

\bibitem{Sabio} J. Sabio, F. Sols, and F. Guinea, Phys. Rev. B \textbf{81},
045428 (2010).

\bibitem{Ashwin} A. Ramasubramaniam, Phys. Rev. B \textbf{86}, 115409 (2012).

\bibitem{Louie} D. Y. Qiu, F. H. da Jornada, and S. G. Louie, Phys. Rev.
Lett. \textbf{111}, 216805 (2013).

\bibitem{Fogler} M.~M. Fogler, L.~V. Butov, and K.~S. Novoselov, Nature
Comm. \textbf{5}, 4555 (2014).

\bibitem{Calman} E. V. Calman, C. J. Dorow, M. M. Fogler, L. V. Butov, S.
Hu, A. Mishchenko, and A. K. Geim, Appl. Phys. Lett. \textbf{108}, 101901
(2016).

\bibitem{Peeters2018} M. Van der Donck and F. M. Peeters, Phys. Rev. B
\textbf{98}, 115104 (2018).

\bibitem{Reichman2013} T. C. Berkelbach, M. S. Hybertsen, and D. R.
Reichman, Phys. Rev. B \textbf{88}, 045318 (2013).

\bibitem{Rubio} P. Cudazzo, I.V. Tokatly, and A. Rubio, Phys. Rev. B \textbf{%
84}, 085406 (2011).

\bibitem{GlazovRMP} G. Wang, A. Chernikov, M. M. Glazov, T. F. Heinz, X.
Marie, T. Amand, and B. Urbaszek, Rev. Mod. Phys. \textbf{90}, 021001 (2018).

\bibitem{DenFuncTheoryPIMC} I. Kyl\"{a}np\"{a}\"{a}, and H.-P. Komsa, Phys.
Rev. B \textbf{92}, 205418 (2015)\emph{.}

\bibitem{BrunettiJP2018} M. N. Brunetti, O. L. Berman, and R. Ya.
Kezerashvili, J. Phys.: Condens. Matter \textbf{30,} 225001 (2018).

\bibitem{WangPhosphorenEx} X. Wang, et al., Nat. Nanotechnol. \textbf{10},
517 (2015).

\bibitem{Rodin} A.~S. Rodin, A. Carvalho, and A.~H. Castro Neto, Phys. Rev.
B \textbf{90}, 075429 (2014).

\bibitem{Prada} E. Prada, J. V. Alvarez, K. L. Narasimha-Acharya, F. J.
Bailen, and J. J. Palacios, Phys. Rev. B \textbf{91}, 245421 (2015).

\bibitem{Hunt22019} R. J. Hunt, M. Szyniszewski, G. I. Prayogo, R. Maezono,
and N. D. Drummond, Phys. Rev. \textbf{98}, 075122 (2018).

\bibitem{RKPhysRevBG2017} O.~L. Berman, G. Gumbs, and R.~Ya. Kezerashvili,
Phys. Rev. B \textbf{96}, 014505 (2017).

\bibitem{Brunetti2019} M. N. Brunetti, O. L. Berman, and R. Ya.
Kezerashvili, Phys. Lett. A \textbf{383}, 482 (2019).

\bibitem{Bartnik} A. C. Bartnik, Al. L. Efros, W.-K. Koh, C. B. Murray, and
F. W. Wise, Phys. Rev. B \textbf{82}, 195313 (2010).

\bibitem{Giblin} J. Giblin, F. Vietmeyer, M. P. McDonald, and M. Kuno, Nano
Lett. \textbf{11}, 3307 (2011).

\bibitem{OgawaTakagahara} T. Ogawa and T. Takagahara, Phys. Rev. B \textbf{44%
}, 8138 (1991).

\bibitem{RK1D} R. Ya. Kezerashvili, Z. Machavariani, B. Beradze, and T.
Tchelidze, Physica E \textbf{109}, 228 (2019).

\bibitem{DasSarma} F. C. Zhang and S. Das Sarma, Phys. Rev. B \textbf{33},
2903 (1986).

\bibitem{Semina} M. A. Semina, R. A. Sergeev, and R. A. Suris,
Semiconductors, \textbf{42}, 1427 (2008).

\bibitem{Bednarek2003} S. Bednarek, B. Szafran, T. Chwiej, and J. Adamowski,
Phys. Rev. B \textbf{68}, 045328 (2003).

\bibitem{Slachmuylders} A. F. Slachmuylders, B. Partoens, W. Magnus, and F.
M. Peeters, J. Phys.: Condens. Matter \textbf{18}, 3951 (2006).

\bibitem{Loudon1} R. Loudon, Am. J. Phys. \textbf{27}, 649 (1959).

\bibitem{Loudon2} R. J. Elliott and R. Loudon, J. Phys. Chem. Solids \textbf{%
8}, 382 (1959).

\bibitem{Loudon3} R. J. Elliott and R. Loudon, J. Phys. Chem. Solids \textbf{%
15}, 196 (1960).

\bibitem{L} M. A. Lampert, Phys. Rev. Lett. \textbf{1,} 450 (1958).

\bibitem{Stebe1987} B. St\'{e}b\'{e}, E. Feddi, and G. Munschy, Phys. Rev. B
\textbf{35}, 4331 (1987).

\bibitem{Stebe1989} B. St\'{e}b\'{e} and A. Ainane, Superlatt. Microstruct.
\textbf{5}, 545 (1989).

\bibitem{QW1} K. Kheng, et al., Phys. Rev. Lett. \textbf{71,} 1752 (1993).

\bibitem{Filikhin} I. N. Filikhin, A. Gal, and V. M. Suslov, Phys. Rev. C
\textbf{68} (2003) 024002.

\bibitem{KezTFSV} R. Ya. Kezerashvili, S. M. Tsiklauri, I. Filikhin, V. M.
Suslov, and B. Vlahovic, J. Phys. G: Nucl. Part. Phys. \textbf{43}, 065104
(2016).

\bibitem{FilKezPLA2018} I. Filikhin, R. Ya. Kezerashvili, and B. Vlahovic,
Phys. Lett. A \textbf{382}, 787 (2018).

\bibitem{QW2} G. Finkelstein, H. Shtrikman, and I. Bar-Joseph, Phys. Rev.
Lett. \textbf{74,} 976 (1995).

\bibitem{QW3} A. J. Shields, M. Pepper, D. A. Ritchie, M. Y. Simmons, and G.
A. C. Jones, Phys.Rev.\textit{\ } B \textbf{51,} 18049 (1995).

\bibitem{QW4} H. Buhmann, L. Mansouri, J. Wang, P. H. Beton, N. Mori, L.
Eaves, M. Henini, and M. Potemski, Phys. Rev. B \textbf{51,} 7969 (1995).

\bibitem{QW5} S. A. Brown, J. F. Young, J. A. Brum, P. Hawrylak, and Z.
Wasilewski, Phys. Rev. B, \textbf{54}, 11082 (1996).

\bibitem{QW6} R. Kaur, A. J. Shields, J. L. Osborne, M. Y. Simmons, D. A.
Ritchie, and M. Pepper, Phys. Stat. Sol. \textbf{178}, 465 (2000).

\bibitem{QW7} V. Huard, R. T. Cox, K. Saminadayar, A. Arnoult, and S.
Tataremko, Phys. Rev. Lett. \textbf{84}, 187 (2000).

\bibitem{Esser1} A. Esser, et al., Phys. Status Solidi (a) \textbf{178}
(2000) 489.

\bibitem{Esser2} A. Esser, et al., Phys. Rev. B \textbf{62}, 8232 (2000).

\bibitem{Bracker} A. S. Bracker, et al., Phys. Rev. B \textbf{72}, 035332
(2005).

\bibitem{Varga1999} J. Usukura, Y. Suzuki, and K. Varga, Phys. Rev. B
\textbf{59}, 5652 (1999).

\bibitem{Varga2000} C. Riva, F.M. Peeters, and K. Varga, Phys. Rev. \textbf{%
61}, 873 (2000).

\bibitem{Ronnow} T.F. R\O nnow, T.G. Pedersen, B. Partoens, K.K. Berthelsen,
Phys. Rev. B \textbf{184}, 035316 (2011).

\bibitem{HHRuan1999} W. Y. Ruan, K. S. Chan, H. P. Ho, R. Q. Zhang, and E.
Y. B. Pun, Phys. Rev. B \textbf{60}, 5714 (1999).

\bibitem{Braun2001} M. Braun and O. I. Kartavtsev, Nucl. Phys. A \textbf{698}%
, 519c (2001).

\bibitem{Braun2004} M. Braun, O. I. Kartavtsev, Phys. Lett. A \textbf{331},
437 (2004).

\bibitem{MoS23Heinz} K. F. Mak, et al., Nat. Mater. \textbf{12,} 207 (2013).

\bibitem{MoSe21 Ross} J. S. Ross, et al., Nat. Comm. \textbf{4,} 1474 (2013).

\bibitem{WSe2 Jones} A. M. Jones, et al., Nat. Nanotechnol. \textbf{8,} 634
(2013).

\bibitem{WSe2Wang} G. Wang, et al., Phys. Rev. B \textbf{90,} 075413 (2014).

\bibitem{MoSe2Singh} A. Singh, et al. Phys. Rev. Lett. \textbf{112,} 21680
(2014)

\bibitem{Liu} C. H. Liu, et al., Phys. Rev. Lett. \textbf{113,} 166801
(2014).

\bibitem{ShangBiexiton} J. Shang, et al., ACS Nano \textbf{9,} 647 (2015).

\bibitem{WS2Plechinger} G. Plechinger, et al. Phys. Status Solidi RRL
\textbf{9,} 457 (2015).

\bibitem{ZhangMS2} Y. Zhang, H. Li, H. Wang, R. Liu, S. Zhang, and Z. Qiu,
ACS Nano \textbf{9,} 8514 (2015).

\bibitem{Christopher2017} J. W. Christopher, B. B. Goldberg, and A. K. Swan,
Sci. Rep. \textbf{7,} 14062 (2017).

\bibitem{CourtadeSemina} E. Courtade, et al., Phys. Rev. B \textbf{96},
085302 (2017).

\bibitem{Bellus} M. Z. Bellus, F. Ceballos, H-Y. Chiu, and H. Zhao, ACS Nano
\textbf{9}, 6459 (2015).

\bibitem{Shang} J. Shang, X. Shen, C. Cong, N. Peimyoo, B. Cao, M.
Eginligil, and T. Yu, ACS Nano \textbf{9}, 647 (2015).

\bibitem{bZu} B. Zhu, H. Zeng, J. Dai, Z. Gong, X. Cui, Proc. Natl. Acad.
Sci. U.S.A. \textbf{111}, 11606 (2014).

\bibitem{Yang} J. Yang, et al., ACS Nano \textbf{9,} 6603 (2015).

\bibitem{Godde} T. Godde, et al., Phys. Rev. B \textbf{94}, 165301 (2016).

\bibitem{WS2Plechinger2016} G. Plechinger, et al., Nat. Comm. \textbf{7},
12715 (2016).

\bibitem{Yang2 2017} J. Yang and Y. Lu, Chin. Phys. B \textbf{26}, 034201
(2017).

\bibitem{Yang3 2016} R. Xu, et al., ACS Nano \textbf{10,} 2046 (2016).

\bibitem{ZhangExTheor} C. Zhang, H.Wang,W. Chan, C.Manolatou, and F. Rana,
Phys. Rev. B \textbf{89}, 205436 (2014).

\bibitem{Falko} M. Szyniszewski, E. Mostaani, N. D. Drummond, and V. I. Fal$%
^{\prime }$ko, Phys. Rev. B \textbf{95}, 081301(R) (2017).

\bibitem{Falko2} E. Mostaani, M. Szyniszewski, C. H. Price, R. Maezono, M.
Danovich, R. J. Hunt, N. D. Drummond, and V. I. Fal$^{\prime }$ko, Phys.
Rev. B \textbf{96,} 075431 (2017).

\bibitem{Ganchev} B. Ganchev, N. Drummond, I. Aleiner, and V. Fal$^{\prime }$%
ko, Phys. Rev. Lett. \textbf{114,} 107401 (2015).

\bibitem{Saxena} K. A. Velizhanin and A. Saxena, Phys. Rev. B \textbf{92},
195305 (2015).

\bibitem{VargaPRB2016} D. W. Kidd, D. K. Zhang, and K. Varga, Phys. Rev. B
\textbf{93}, 125423 (2016).

\bibitem{BerkelbachDifMonteCarlo} M. Z. Mayers, T. C. Berkelbach, M. S.
Hybertsen, and D. R. Reichman, Phys. Rev. B \textbf{92}, 161404 (2015).

\bibitem{Thilagam1997} A. Thilagam, Phys. Rev. B \textbf{55}, 7804 (1997).

\bibitem{ZareniaPeeters2017} M. Van der Donck, M. Zarenia, and F. M.
Peeters, Phys. Rev. B \textbf{96}, 035131 (2017).

\bibitem{KezFBS2017} R.~Ya. Kezerashvili and S. M. Tsiklauri, Few-Body Syst.
\textbf{58}, 18 (2017).

\bibitem{KezFil2018} I. Filikhin, R. Ya. Kezerashvili, S. M. Tsiklauri, and
B. Vlahovic, Nanotechnol. \textbf{29}, 124002 (2018).

\bibitem{TimeDepdensity matrix functional theory} A. Ramirez-Torres, V.
Turkowski, and T.S. Rahman, Phys. Rev. B \textbf{90}, 085419 (2014).

\bibitem{KezIJMPB2014} O.~L. Berman, R.~Ya. Kezerashvili, and S. M.
Tsiklauri, Int. J. Mod. Phys. B \textbf{28}, 1450064 (2014).

\bibitem{Bondarev2017} I. V. Bondarev and M. R. Vladimirova, Phys. Rev. B
\textbf{97}, 165419 (2018).

\bibitem{Kimgroup2018} L. A. Jauregui, et al., arXiv:1812.08691
[cond-mat.mes-hall] (2018).

\bibitem{Drummond2018} O. Witham, R. J. Hunt, and N. D. Drummond, Phys. Rev.
B \textbf{97}, 075424 (2018).

\bibitem{QW1 Baars} T. Baars, W. Braun, M. Bayer, and A. Forchel, Phys. Rev.
B \textbf{58}, R1750 (1998).

\bibitem{Zimmermann} A. Esser, R. Zimmermann, and E. Runge, Phys. Stat. Sol.
B \textbf{227}, 317 (2001).

\bibitem{Tsu2001} T. Tsuchiya, Int. J. Mod. Phys. B \textbf{15}, 3985 (2001).

\bibitem{1DEXP} A. Crottini, J. L. Staehli, B. Deveaud, X. L.Wang, and M.
Ogura, Solid State Comm.. \textbf{121}, 401 (2002).

\bibitem{ExQW} T. Otterburg, et al., Phys. Rev. B \textbf{71}, 033301 (2005).

\bibitem{PeetersPRB05} B. Szafran, T. Chwiej, F. M. Peeters, S. Bednarek,
and J. Adamowski, Phys. Rev. B \textbf{71}, 235305 (2005).

\bibitem{PeetersFewBody2006} F. M. Peeters, B. Szafran, T. Chwiej, S.
Bednarek, and J. Adamowski, Few-Body Syst. \textbf{38}, 121 (2006).

\bibitem{PeetersPRB2008} Y. Sidor, B. Partoens, and F. M. Peeters, Phys.
Rev. B\ \textbf{77}, 205413 (2008).

\bibitem{Piermarocchi} M. J. A. Schuetz, M. G.Moore, and C. Piermarocchi,
Nat. Phys. \textbf{6}, 919 (2010).

\bibitem{QW 2012 Phys Lett} L. X. Zhai, Y. Wang, and J. J. Liu, Phys. Lett.
A \textbf{376}, 1866 (2012).

\bibitem{Pedersen} T. G. Pedersen, Phys. Rev. B \textbf{67}, 073401 (2003).

\bibitem{Pedersen2} T. G. Pedersen, K. Pedersen, H. D. Cornean, and P.
Duclos, Nano Lett. \textbf{5}, 291 (2005).

\bibitem{Kammerlander} D. Kammerlander, D. Prezzi, G. Goldoni, E. Molinari,
and U. Hohenester, Phys. Rev. Lett. \textbf{99}, 126806 (2007); Physica. E
\textbf{40}, 1997 (2008).

\bibitem{Pedersen3} T. F. R\o nnow, T. G. Pedersen, and H. D. Cornean, Phys.
Rev. B \textbf{81}, 205446 (2010).

\bibitem{DifMC} T. F. R\o nnow, T. G. Pedersen, B. Partoens, and K. K.
Berthelsen, Phys. Rev. B \textbf{84}, 035316 (2011).

\bibitem{Matsunaga} R. Matsunaga, K. Matsuda, and Y. Kanemitsu, Phys. Rev.
Lett. \textbf{106},\ 037404 (2011).

\bibitem{Santos} S. M. Santos, B. Yuma, S. Berciaud, J. Shaver, M. Gallart,
P. Gilliot, et al., Phys. Rev. Lett. \textbf{107}, 187401 (2011).

\bibitem{Bondarev} I. V. Bondarev, Phys. Rev. B \textbf{83}, 153409 (2011).

\bibitem{Watanabe} K. Watanabe and K. Asano, Phys. Rev. B \textbf{83},
115406 (2011); Phys. Rev. B \textbf{85}, 035416 (2012).

\bibitem{Pedersen4} T. F. R\o nnow, T. G. Pedersen, and B. Partoens, Phys.
Rev. B \textbf{85}, 045412 (2012).

\bibitem{ColombierPRL} L. Colombier, J. Selles, E. Rousseau, J. S. Lauret,
F. Vialla, C. Voisin, and G. Cassabois, Phys. Rev. Lett. \textbf{109},
197402 (2012).

\bibitem{Yuma} B. Yuma, S. Berciaud, J. Besbas, J. Shaver, S. Santos, S.
Ghosh, et al., Phys. Rev. B \textbf{87}, 205412 (2013).

\bibitem{Bondarev2} I. V. Bondarev, Phys. Rev. B \textbf{90}, 245430 (2014).

\bibitem{Bondarev3} I. V. Bondarev, Mod. Phys. Lett. B \textbf{30}, 1630006
(2016).

\bibitem{Slachmuylders1} A. F. Slachmuylders, B. Partoens, W. Magnus, and F.
M. Peeters,\textbf{\ }Phys. Rev. B \textbf{76}, 075405 (2007).

\bibitem{Slachmuylders2} A. F. Slachmuylders, B. Partoens, W. Magnus, and F.
M. Peeters, Physica E \textbf{40}, 2166 (2008).

\bibitem{Deilmann2016} T. Deilmann, M. Dr\"{u}ppel, and M. Rohlfing, Phys.
Rev. Lett. \textbf{116}, 196804 (2016).

\bibitem{Lozovik1D} L. N. Ivanov, Yu. E. Lozovik, and D. R. Musin, J. Phys.
C: Solid State Phys. \textbf{11}, 2527 (1978).

\bibitem{MillerQW1982} R. C. Miller, et al., Phys. Rev. B \textbf{25}, 6545
(1982).

\bibitem{Bressanini1998} D. Bressanini, M. Mella, and G. Morosi, Phys. Rev.
A \textbf{55}, 200, (1997); Phys. Rev. A \textbf{57}, 4956 (1998).

\bibitem{MaiBiexiton} C. Mai et al., Nano Lett. \textbf{14}, 202 (2014).

\bibitem{Reichman2015} Y. You, X.-X. Zhang, T. C. Berkelbach, M. S.
Hybertsen, D. R. Reichman, and T. F. Heinz, Nat. Phys. \textbf{11}, 477
(2015).

\bibitem{SieBiexitonMoS2} E. J. Sie, A. J. Frenzel, Y.-H. Lee, J. Kong, and
N. Gedik, Phys. Rev. B \textbf{92}, 125417 (2015).

\bibitem{MoSe2Hao} K. Hao et al., Nat. Comm. \textbf{8},15552 (2017).

\bibitem{VargaNano2015} D. K. Zhang, D. W. Kidd, and K. Varga, Nano Lett.
\textbf{15}, 7002 (2015).

\bibitem{Thilagam2014} A. Thilagam, J. Appl. Phys \textbf{116}, 053523
(2014).

\bibitem{TuanNewRKPot} D. V. Tuan, M. Yang, and H. Dery, Phys. Rev. B
\textbf{98}, 125308 (2018).

\bibitem{Chen2018} S.-Y. Chen, T. Goldstein, T. Taniguchi, K. Watanabe, and
J. Yan, Nat. Comm. \textbf{9}, 3717 (2018).

\bibitem{Waldecker2018} Z. Ye, L. Waldecker, E.Y. Ma, D. Rhodes, A. Antony,
B. Kim, et al., Nat. Comm. \textbf{9}, 3718 (2018).

\bibitem{Li2018} Z. Li, T. Wang, Z. Lu, C. Jin , Y. Chen, Y. Meng, et al.,
Nat. Comm. \textbf{9}, 3719 (2018).

\bibitem{Barbone2018} M. Barbone, A. R.-P. Montblanch, D. M. Kara, C.
Palacios-Berraquero, A. R. Cadore, D. De Fazio, et al., Nat. Comm. \textbf{9}%
, 3721 (2018).

\bibitem{Steinhoff2018} A. Steinhoff, et al., Nat. Phys. \textbf{14}, 1199
(2018).

\bibitem{Fogler2008} A. D. Meyertholen and M. M. Fogler, Phys. Rev. B
\textbf{78}, 235307 (2008).

\bibitem{Pack1970} R. T. Pack and J. O. Hirschfelder, J. Chem. Phys. \textbf{%
52}, 521 (1970).

\bibitem{Tan2005} M. Y. J. Tan, N. D. Drummond, and R. J. Needs, Phys. Rev.
B \textbf{71}, 033303 (2005).

\bibitem{ReichmanPhosforos} A. Chaves, M. Z. Mayers, F. M. Peeters, and D.
R. Reichman, Phys. Rev. B \textbf{93}, 115314 (2016).

\bibitem{Crottini1D} A. Crottinia, J.L. Staehlia, B. Deveauda, X.L. Wang, M.
Ogura, Solid State Comm. \textbf{121}, 401 (2002).

\bibitem{Baer2013} R. Baer and E. Rabani, J. Chem. Phys \textbf{138}, 051102
(2013).

\bibitem{Herring1964} C. Herring and M. Flicker, Phys. Rev. \textbf{134},
A362 (1964).

\bibitem{Timofeyuk} N. K. Timofeyuk and D. Baye, Few-Body Syst. \textbf{58},
157 (2017).

\bibitem{Kim2016} P. Ajayan, P. Kim, and K. Banerjee, Phys. Today \textbf{69}%
, 9, 38 (2016).

\bibitem{Kogar2017} A. Kogar, et al., Science \textbf{358}, 1314 (2017).

\bibitem{Cotlet2016} O. Cotle\c{t}, S. Zeytino{g}lu, M. Sigrist, E. Demler,
and A. Imamo{g}lu, Phys. Rev. B \textbf{93}, 054510 (2016).

\bibitem{ThygesenReview2017} K. S. Thygesen, 2D Mater. \textbf{4,} 022004
(2017).
\end{thebibliography}
\end{document}